\documentclass{aa}  

\usepackage{graphicx}
\usepackage{txfonts}
\usepackage{comment}
\usepackage{amssymb}

\usepackage[normalem]{ulem}
\usepackage{amsmath}
\usepackage{cases}
\usepackage{lineno}
\usepackage[colorlinks=true, linkcolor=blue, citecolor=blue, urlcolor=blue]{hyperref}
\newcommand{\unit}[1]{\boldsymbol{\rm{\hat{#1}}}}

\newcommand{\obs}{_{\rm obs}}

\newcommand{\Pmax}{P_{\max}}

\begin{document}

   \title{Afterglow Linear Polarization Signatures from Steep GRB Jets: Implications for Orphan Afterglows}


   \author{G. Birenbaum
          \inst{1,2}\thanks{Corresponding author:birenbaumgal@gmail.com}
          \and
         J. Granot\inst{1,2,3}
          \and
          P. Beniamini\inst{1,2,3}
          }

   \institute{Astrophysics Research Center of the Open University (ARCO), The Open University of Israel, P.O. Box 808, Ra’anana 4353701, Israel
         \and
             Department of Natural Sciences, The Open University of Israel, P.O. Box 808, Ra’anana 4353701, Israel
        \and
            Department of Physics, The George Washington University 
Washington, DC 20052, USA
             }


 
  \abstract
{Gamma-Ray Bursts (GRBs) are the strongest explosions in the Universe, and are powered by initially ultra-relativistic jets. The angular profile of GRB jets encodes important information about their launching and propagation near the central source, and can be probed through their afterglow emission. 
Detailed analysis of the multi-wavelength afterglow light curves of recent GRBs indeed shows evidence for an extended angular structure beyond the jet's narrow core. The afterglow emission is determined by the jet angular structure, our viewing angle, and the magnetic field structure behind the shock, often leading to degeneracies when considering the light curves and broad-band spectrum alone.
Such degeneracies can be lifted with joint modeling of the afterglow light curves and polarization. 
In this work we study the evolution of the afterglow linear polarization and flux density from steep, core-dominated GRB jets, where most of their energy resides within a narrow core. We explore the dependence of the light and polarization curves on the viewing angle, jet angular energy structure and magnetic field configuration, and provide an analytical approximation for the peak polarization level, which occurs at a time close to that of a break in the light curve.
Finally, we demonstrate how our results can be used to determine the nature of orphan GRB afterglows, distinguishing between a quasi-spherical "dirty fireball" and a steep jets viewed far off-axis and apply them on the Zwicky Transient Facility (ZTF) detected orphan afterglow candidate AT2021lfa. }

   \maketitle
%

\section{Introduction}
\label{sec:intro}

Gamma-ray bursts (GRBs) are violent astrophysical events associated with the launch of relativistic jets, that originate in newly-born accreting compact objects \citep{Eichler1988,MacFadyen1999}. Such systems are not spherically symmetrical and the brief prompt $\gamma$-ray emission can be observed only when the observer is located within, or very close to, the jet opening angle \citep[e.g.][]{2019MNRAS.482.5430B}. GRBs are followed up with long-lasting emission, termed the "afterglow". This multi-wavelength component is associated with the formation of a relativistic shock, due to the interaction of the relativistic outflow with the ambient medium. The shock accelerates charged particles that radiate linearly polarized synchrotron radiation under the influence of shock-generated magnetic fields \citep[e.g.][]{Paczynski1993,1994ApJ...422..248K,1997ApJ...490..772K,1997ApJ...489L..33W,1997ApJ...485L...5W,Sari1998,1998ApJ...499..301M, Gruzinov1999, Ghisellini1999, Sari1999Pol}. Earlier works on GRBs assumed the jet has a simple "top hat" angular structure, in which the energy and Lorentz factor are uniform within a narrow core opening angle and drop sharply beyond it. Such structure was shown to explain the appearance of achromatic steepenings in the afterglow light curve, caused by the deceleration of the flow and the Lorentz factor reaching a value comparable to the inverse of half of the core opening angle. Following this stage, there are no new areas of the jet that are unveiled to the observer as it decelerates, leading to a geometrical "jet break" in the afterglow light curve \citep{Rhoads1997,Rhoads1999, Sari1999JetBreak}. However, detailed afterglow light curve analysis of recent events shows evidence for the existence of an extended angular structure of the jet beyond its core, most notably, GW 170817 \citep{GG18,Lazzati2018,Mooley2018,Ghirlanda2019,Gill+19,Troja2019,Govreen-Segal2023}. In some cases, different jet structures may reproduce the same multi-wavelength light curves, making the jet structure difficult to constrain based on light curve fits alone. In the case of GRB 221009A, the multi-wavelength afterglow light curves can be reproduced with two different jet structures \citep{O'Connor2023,Gill2023}. In \citet{Birenbaum2024}, we show that while the existing polarization limits on the X-ray and optical afterglow \citep{Negro2023} agree with both models, an earlier measurement could have differentiated between the models, making this observable a unique tool for probing the angular structure of the jet.

Numerical simulations suggest that the jet develops an extended angular structure during its propagation within the progenitor system \citep{Gill+19, Gottlieb2021, Govreen-Segal2024} which can be merger ejecta in the case of a short GRB \citep{Eichler1988,Narayan1992,Berger2013,Tanvir2013,Abbott2017,Mooley2018} or stellar envelope in the case of a long GRB \citep{MacFadyen1999,Galama1998,Woosley2006,Metzger2011}. The structure the jet emerges with reflects the processes it underwent before breaking out, as well as jet production conditions, making the jet structure an important quantity to constrain. Steep jets, which hold most of the energy within the jet core, can be formed when the jet is weakly magnetized, which acts to reduce the amount of mixing between the light jet material with that of the heavy confining medium via suppression of hydrodynamical instabilities \citep{Gottlieb2020,Beniamini2020}. Such structure can also be consistent with most afterglow observations.

In recent years, optical transients with afterglow-like temporal evolution and no associated GRB, have been detected using the Zwicky Transient Facility (ZTF). The origin of these events is under debate \citep{Lipunov2022,Ho2022,Li2024}. At the time of writing, nine such events were found using ZTF, three of which without retroactively associated prompt $\gamma$-ray counterpart \citep{Ho2020, Ho2022, Andreoni2021, Perley2025}. 
One of these events, AT2021lfa, is detected in optical, radio and X-ray.
The optical light curve features clearly rising flux which peaks and then declines as a power-law. While the declining phase of the transient was initially captured by ZTF, the preceding rising phase was serendipitously found in archival MASTER-OAFA data \citep{Lipunov2021,Lipunov2022}. Comparison of the optical light curve with known on-axis GRB afterglow light curves, conducted by \citet{Lipunov2022}, shows similarities which motivate this transient's relation to GRBs. The authors fit top hat jet optical afterglow models to the light curve and vary the viewing angle, resulting in the on-axis model fitting the data better than the off-axis one. They concluded that this ``orphan afterglow" candidate is the result of a ``dirty fireball" with baryon-contaminated ejecta that fails to produce a GRB. However, the off-axis afterglow scenario the authors fit does not include the possibility of a steep structured jet, which can also exhibit a slow rise in flux, a peak and a power-law decline in time \citep{Granot2002,Kumar-Granot2003,Nakar2020,Gill2020,Beniamini2022}. This possibility was considered in recent work that revisits the multi-wavelength observations of AT2021lfa and manages to fit the optical light curve with both on-axis jet and off-axis jet models using the \texttt{afterglowpy} tool \citep{Ryan2020, Li2024}. The on-axis jet fit can be obtained using both top hat and Gaussian jets. The off-axis jet fits consider the structured jet scenario for the optical afterglow and are able to fit both power-law and Gaussian jets. 

While the light curves of these events fail to pinpoint the exact geometry of the system and whether it is on- or off-axis, such events are expected to differ greatly in polarized light \citep{Granot2002, Rossi2004, Gill2020, Teboul2021, Birenbaum2024}. In addition, the detection of linear polarization can confirm the synchrotron source of the emission \citep[e.g.][]{Gruzinov1999,Ghisellini1999,Sari1999Pol}.
Measuring this quantity along with the light curve in the optical can indicate whether the system originates in a dirty fireball or in a structured off-axis jet. 

In this paper we characterize the polarization signature from steep jets, starting with structure dependence of on- and off-axis jets and then expanding the off-axis jet analysis to viewing angle effects and the impact of the magnetic field structure behind the shock on the observed polarization and flux. Building on the conclusions drawn from these models, we offer an analytical approximation for the dependence of the polarization peak on the geometrical parameters of the system. Finally, we demonstrate the ability of polarization modeling and measurements in discerning between dirty fireballs and off-axis structured jets on the orphan afterglow candidate AT2021lfa, extending the work done by \citet{Li2024} to the polarized regime.
\section{Methods}
\label{sec:Methods}
\begin{figure}
	\includegraphics[width=\columnwidth]{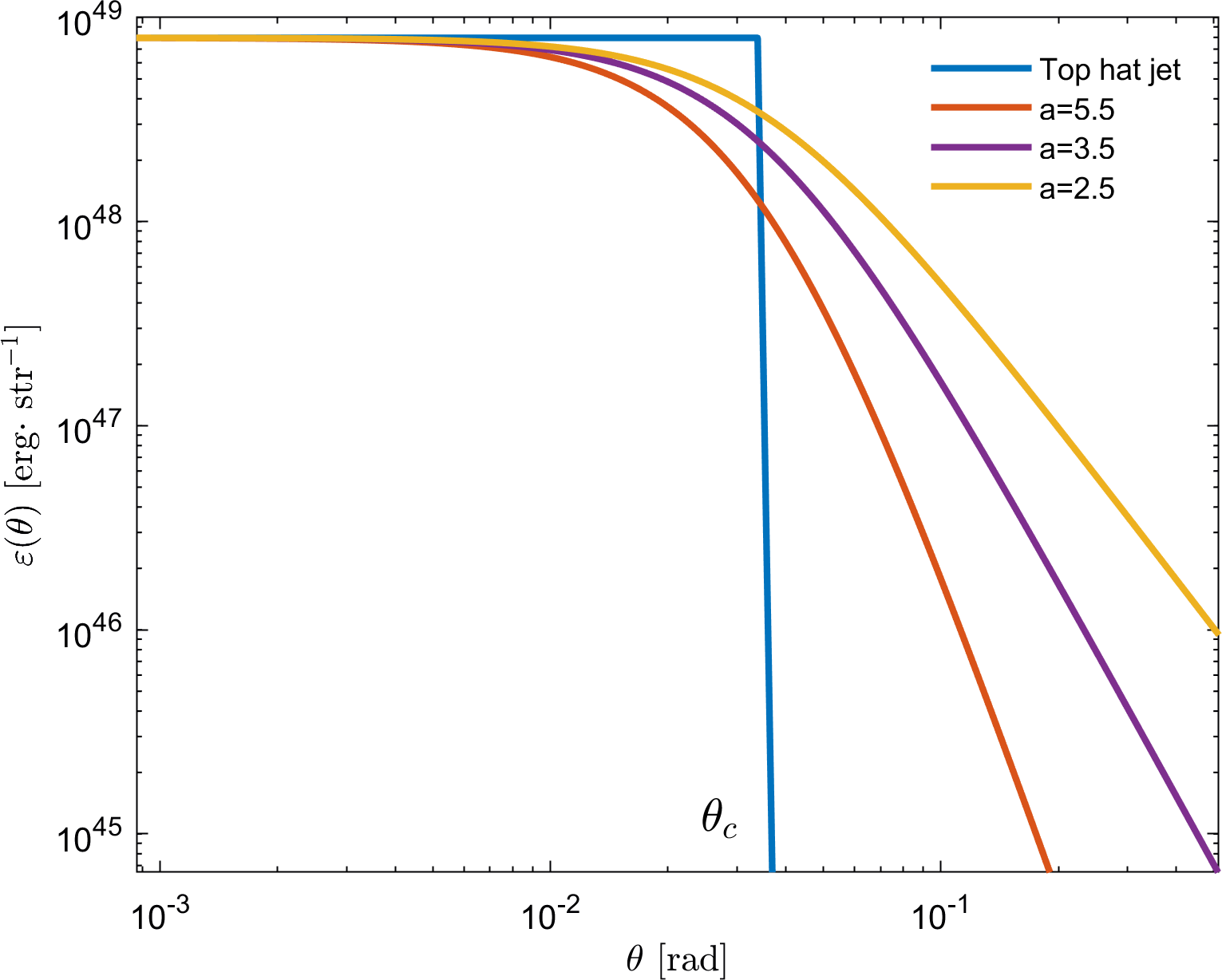}
    \caption{The jet angular energy structures considered in this work with core opening angle $\theta_{\text{c}}=2^{\circ}$. In blue lines we present the top hat jet model which acts as a step function in energy. The other models behave according to the power-law jet model with varying steep slopes with $a>2$ (see \S\,\ref{sec:Methods} and B24).}
 \label{fig:EnergyPerUnitSolidAngleExample}
\end{figure}
The calculations presented in this work follow the formulas presented in the methods section of \citet[][B24 hereafter]{Birenbaum2024}. The basic model assumed in 
B24 and this work features a 2D axis-symmetric relativistic shock propagating
into a cold ambient medium with a power-law rest-mass density profile $\rho\propto R^{-k}$. Following \citet{GG18}, the jet dynamics are assumed to be locally spherical (neglecting lateral dynamics) and the emission is calculated from a 2D surface associated with the afterglow shock front. The jet angular profile is expressed in terms of the distribution of the isotropic equivalent kinetic energy and initial Lorentz factor, which are described by $E_{\rm k,iso}=E_c\Theta^{-a}$ (with $a>2$ corresponding to a steep jet, see Fig. \ref{fig:EnergyPerUnitSolidAngleExample}) and $\Gamma_0-1=(\Gamma_c-1)\Theta^{-b}$ where $\Theta=[1+(\theta/\theta_c)^2]^{1/2}$.
The subscript c denotes properties of the jet core.
The shock surface is divided into angular cells around the jet symmetry axis, defined by $\theta$. The emitting region right behind the shock radiates synchrotron emission under the influence of a shock generated magnetic field whose comoving direction $\unit{B}'$ and corresponding magnitude $B'$ in each cell are drawn from a probability distribution set by the magnetic field stretching factor
$\xi$ (for a more detailed explanation, see Appendices A and B of B24). 

Following these initial calculations, we proceed to calculate the Stokes parameters $I_{\nu},Q_{\nu},U_{\nu}$ from each cell and integrate over their contributions in order to evaluate the overall observed linear polarization and flux.

Full details of the calculation are presented in B24.

\section{Results}
\label{sec:Results}

As shown in B24
and \cite{Rossi2004}, the angular structure of the jet has a profound impact on the polarization signature of GRB afterglows.
The shallow jet case ($a<2$) was studied in detail in B24. Here we extend and complete that study by
exploring the parameter regime of steep jets ($a>2$), which are often referred to as core-dominated jets, since most of their energy resides in their narrow cores. In this work
the same afterglow parameters
chosen in B24
are selected so that the observed frequency does not cross any of the critical synchrotron frequencies (see Table \ref{tab:ParameterSpace}). In the sections below, results are shown for both on-axis jets, viewed from within their core opening angle (with normalized viewing angles $q\equiv \theta\obs/\theta_c<1$) and off-axis jets, (with $q>1$). The cases considered in this section are set with a uniform initial Lorentz factor profile ($b=0$).

\begin{table}[h!]
\caption{Parameter space and afterglow model for the cases  considered in \S \ref{sec:Results}.}
\label{tab:ParameterSpace}
\centering
\renewcommand{\arraystretch}{1.4} 
\begin{tabular}{l c c c}
\hline\hline
 Parameter & Value \\
 \hline
 $\theta_c$    &$2^{\circ}$\\
 $\Gamma_c$  & 250\\
 $a=-\frac{d\log E_{\rm k,iso}}{d\log\theta}|_{\theta>\theta_c}$ 
 &   2.5, 3.5, 4.5, 5.5, Top-hat jet \\
 $b=-\frac{d\log(\Gamma_0-1)}{d\log\theta}|_{\theta>\theta_c}$ 
 & 0 \\
 $E_{\text{c}}$ [erg]& $10^{50}$   \\
 $n_{\text{ISM}}\ [\textrm{cm}^{-3}]$&  1  \\
 $k=-\frac{d\rho}{d\log r}$

 & 0 
 \\
 $\nu_{\text{obs}}$ [Hz]&  $10^{15}$ (PLS G)\\
  $p$&  $2.5$  \\
 $\epsilon_\text{e}$   & 0.1   \\
 $\epsilon_\text{B}$&  0.005\\
  $\chi_\text{e}$&  1\\
 $d_{\text{L}}$ [cm]&  $ 10^{28}$\\
 $z$&  $ 0.54$\\
\hline
\end{tabular}
\end{table}

\subsection{Viewing angle and jet structure }
\subsubsection{Jets viewed on-axis ($0<q<1$)}
\label{subsec:OnAxisJets}

\begin{figure}
	\includegraphics[width=\columnwidth]{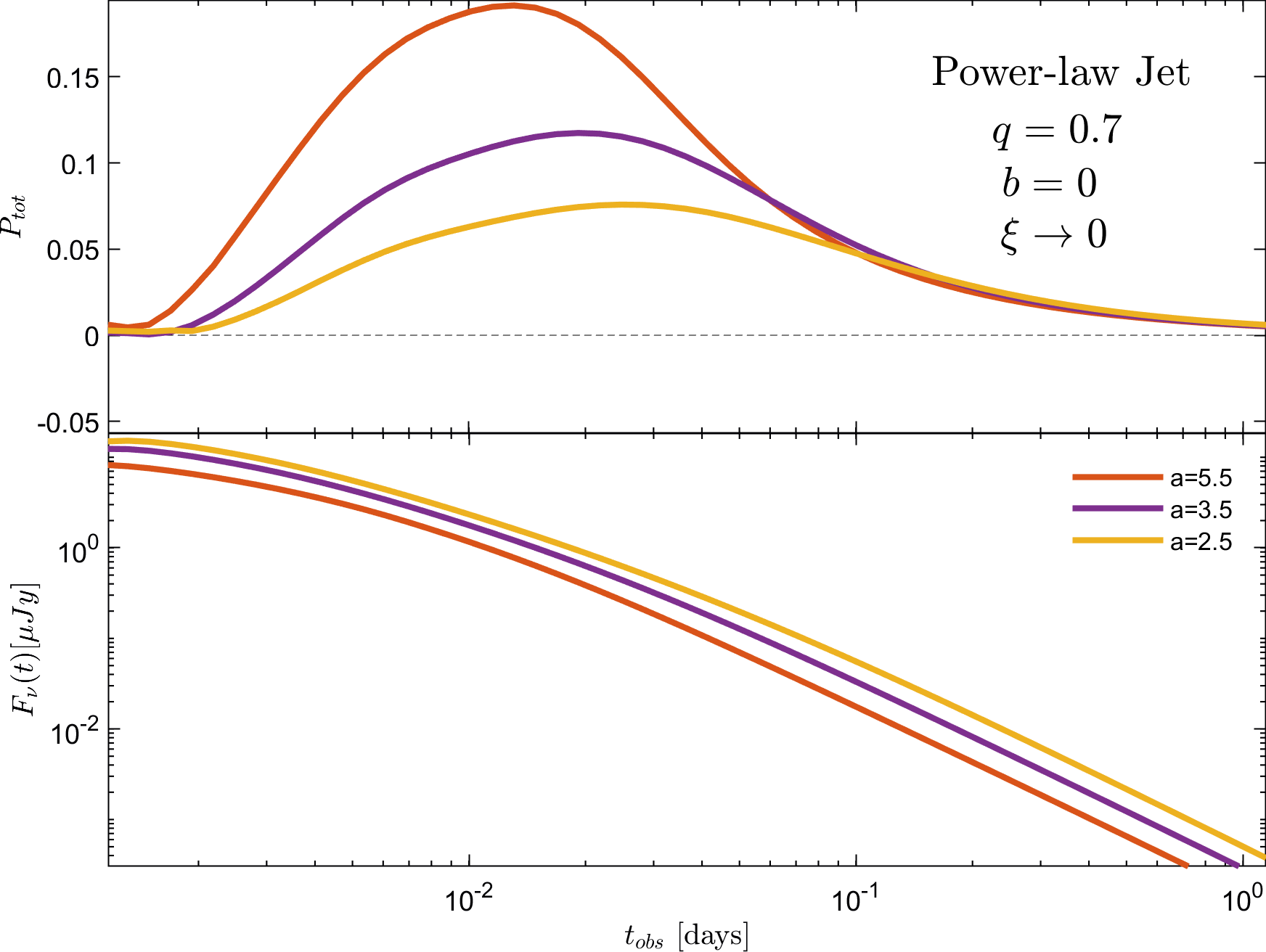}
    \caption{Observed polarization degree (\textit{upper panel}) and flux (\textit{lower panel}) as function of time for the power-law jet model at observed frequency $\nu=10^{15}$ Hz (PLS G) with a flat initial Lorentz factor distribution ($b=0$) and a random magnetic field structure, confined to the shock plane ($\xi\rightarrow 0$). The various values of $a$ represent different slopes for the power-law jet wings and the observer is located at $q=0.7$. 
}
    \label{fig:q<1PDAllModels}
\end{figure}
We start by exploring the polarization and light curves of steep jets from on-axis jets, set with $q\equiv\theta\obs/\theta_c=0.7$. This parameter regime is explored within the context of the Power-law Jet structure, described in Eqs. (2)-(4) of B24 (see Fig. \ref{fig:EnergyPerUnitSolidAngleExample}). 
In Fig. \ref{fig:q<1PDAllModels}, we present the polarization curves (upper panel) and light curves (lower panel) for models with a constant normalized viewing angle $q=0.7$, magnetic field structure that is random within the plane of the shock ($\xi\to0$) and varying angular steep jet structures ($a>2$).

The polarization curves of the various angular energy structures all exhibit a single polarization peak and no rotation of the polarization angle (polarization degree remains positive throughout). Such behavior of the observed polarization differs from that of top hat jets and broken power-law jets, which 
feature a $90^\circ$ rotation of the polarization angle 
that is manifested in our formalism as a change in the sign of the polarization degree (\citealt{Sari1999Pol,Ghisellini1999, Granot2003a,Rossi2004,Shimoda2020,Birenbaum2021}; B24). This new behavior of on-axis jets stems from the use of the Power-Law Jet model, which features a smooth transition between the jet core and the extend jet structure and can eliminate the polarization angle rotation altogether. This important feature will be studied in detail in future work (Birenbaum et al. in prep). 

While the shape of the polarization curve remains similar, its magnitude varies and grows as the structure becomes steeper with growing values of $a$. This happens due to increased asymmetry in the visible part of the emitting region as the contribution of the jet core dominates the polarized emission with less contribution from the extended jet structure. 

The light curves of these models, varying only by the power-law index of the jet angular energy profile, show similar shapes while varying slightly in terms of magnitude, where shallower structures, with lower values of $a$, feature higher observed flux as they contain more energy. 
While the light curves are similar to one another, each structure has its own unique polarization signature, and with sufficient measurements and careful modeling, the jet structure
can be determined.

\subsubsection{Jets viewed off-axis ($q\ge 1$)}
\label{subsection:OffAxisJets}

\begin{figure*}
  \centering
  {\includegraphics[width=\columnwidth]{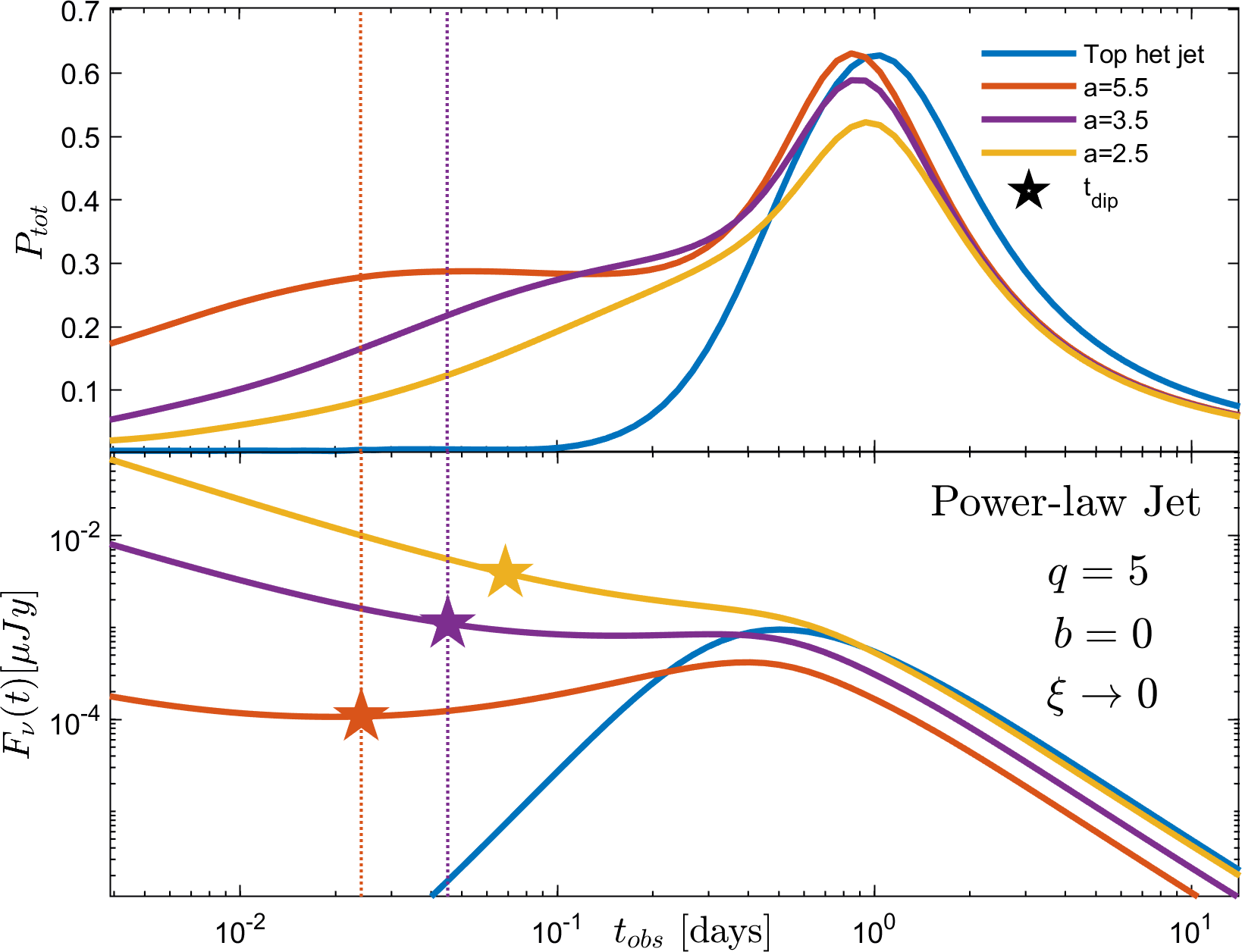}} 
  {\includegraphics[width=\columnwidth]{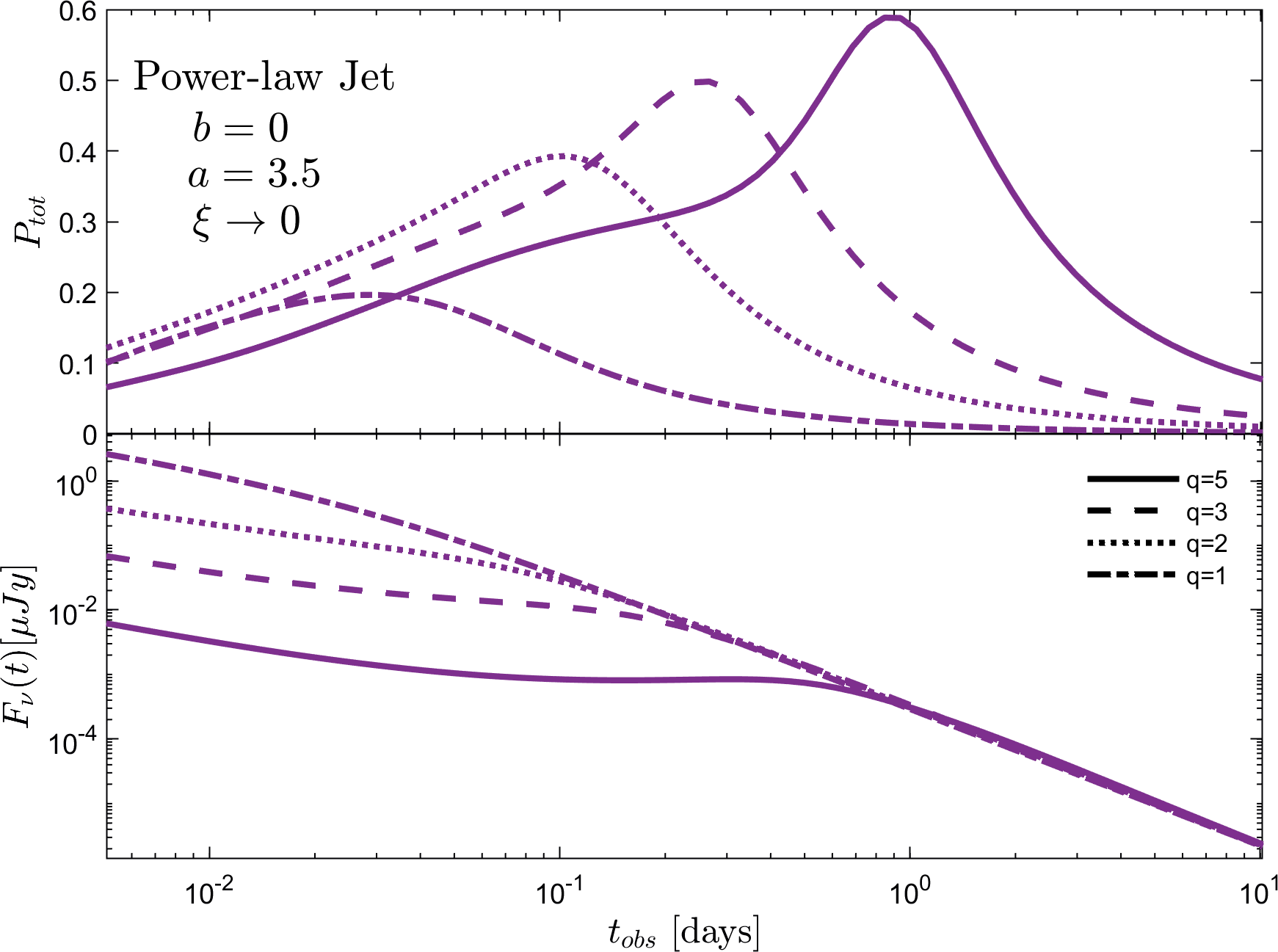}} 

  \caption{Observed polarization degree (\textit{upper panels}) and flux (\textit{lower panels}) at observer frequency $\nu=10^{15}$ Hz 
  (PLS G) with a flat distribution of initial Lorentz factor ($b=0$) and a random magnetic field  structure confined to the face of the shock ($\xi\rightarrow 0$) for off-axis jets. The direction of the polarization vector remains constant throughout the temporal evolution. \textbf{\textit{Left panel}}: We set the viewing angle to be constant for an off-axis jet with $q=5$ with varying values of $a$. The polarization degree reduces with $a$ and its peak coincides with a light curve break. We mark with a star and vertical dashed lines the time of a light curve dip, expected when the emission transitions from angular structure dominated to core dominated in the Blandford-Mckee deceleration regime \citep{Beniamini2020}. At this time, the onset on a plateau phase can be seen in the polarization curve. \textbf{\textit{Right panel}}: The structure of the jet is held constant with $a=3.5$ and the off-axis viewing angle is changed. As the observer line of sight approaches the jet symmetry axis with reducing values of $q$, the polarization peak becomes lower and occurs at earlier times.}
 
\label{fig:P2Together}
\end{figure*}
\begin{figure}
	\includegraphics[width=\columnwidth]{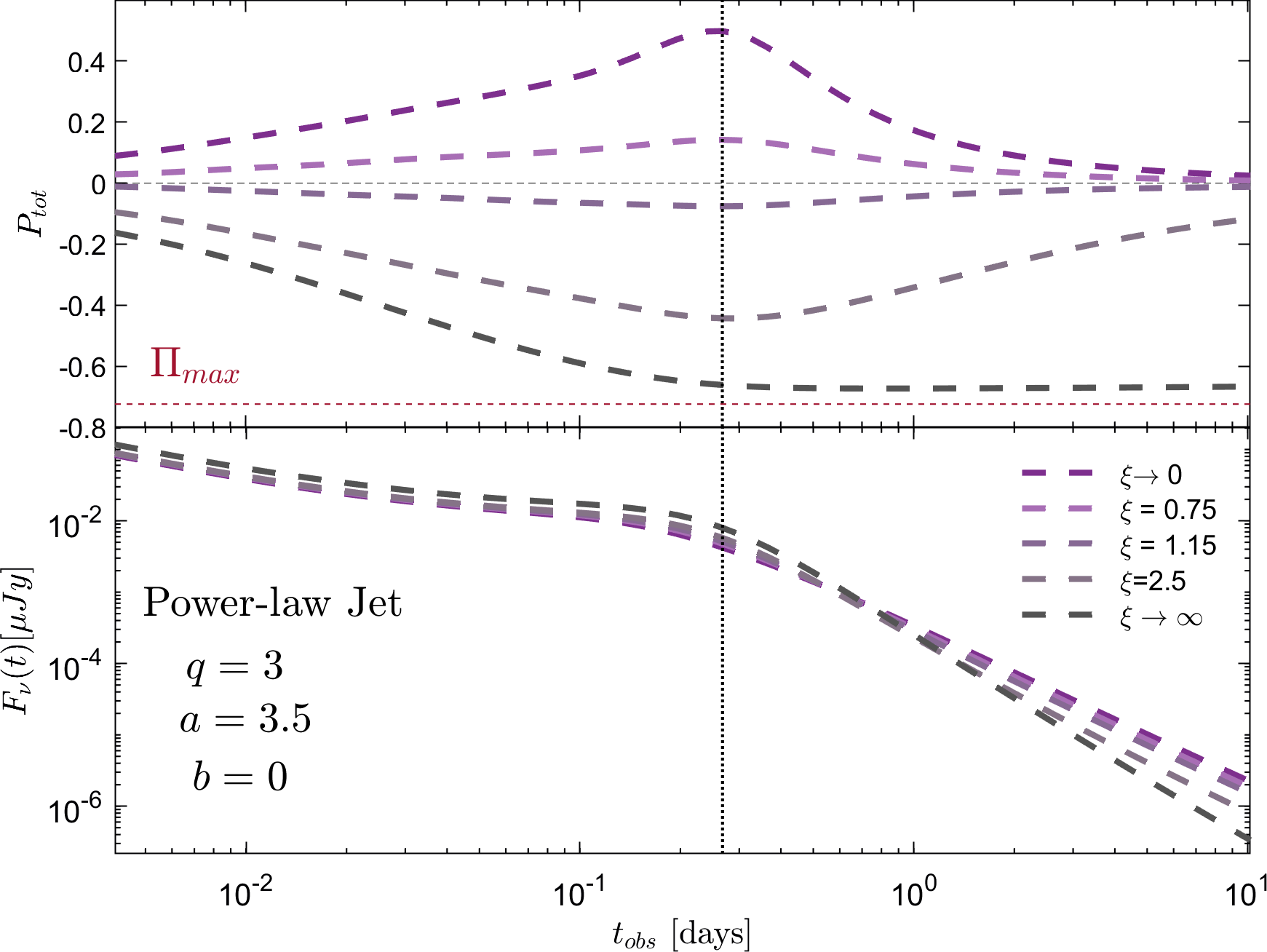}
    \caption{Observed polarization degree (\textit{upper panel}) and flux (\textit{lower panel}) at observer frequency $\nu=10^{15}$ Hz (PLS G) with a flat distribution of initial Lorentz factor ($b=0$), smooth power law energy profile with $a=3.5$, observed with $q=3$. The structure of the magnetic field changes with the value of $\xi$. 
    The direction of the polarization vector changes by $90^{\circ}$ once the stretching factor $\xi$ crosses $1$, manifesting as a change in the sign of the polarization degree. The polarization peak happens at about the same time for all curves and is close to the break time of the light curve. }
\label{fig:q>PDFAllXi}
\end{figure}
In this section, we describe the polarization signature of steep jets ($a>2$) viewed off-axis ($q\geq1$). This setup is studied using the smooth Power-law Jet configuration, as it provides a smooth transition between the jet core and its extended structure (i.e. power-law wings, see Fig. \ref{fig:EnergyPerUnitSolidAngleExample}).

We start by exploring the dependence of the light and polarization curves on the angular energy profile 
of the jet by varying the value of $a$, and setting the normalized viewing angle to be constant with $q=5$ (see left panel of Fig.~\ref{fig:P2Together}). The magnetic field structure behind the shock is set to be random in the plane of the shock with $\xi\rightarrow 0$. The case of a misaligned top hat jet is shown in blue solid lines for comparison purposes.
Similarly to the behavior shown in \S \ref{subsec:OnAxisJets}, the height of the polarization peak increases with the steepness of the jet, encapsulated by rising values of $a$. We also note that the polarization peak is associated with a break in the light curve, similar to the behavior seen in shallow jets and on-axis steep jets (see B24 and Fig. \ref{fig:q<1PDAllModels}).
As emission from the core starts to reach the observer, asymmetry in the visible region reaches its peak and so does the polarization. Following the revelation of the most energetic part of the jet to the observer, the light curve experiences a steepening. This effect is purely geometrical and is therefore largely achromatic.\footnote{A small degree of chromaticity is still possible between different power-law segments of the spectrum, which have a different dependence on $E_{\rm k,iso}$, leading to somewhat different weights for different parts of the visible region.} 

The height of the polarization peak for the shallowest structure considered in this work with $a=2.5$ (left panel of Fig.~\ref{fig:P2Together},
yellow solid curves) is the lowest one. This corresponds to the conclusion drawn in B24 
and in 
\S\,\ref{subsec:OnAxisJets} that shallower jets produce lower levels of polarization due to increased levels of symmetry around the line of sight. The polarization peak of the steepest structure considered in this work, with $a=5.5$ (left panel of Fig.~\ref{fig:P2Together}, 
red solid curves) converges to the polarization peak of the top hat jet structure (solid blue curves) at the same misalignment, indicating maximal asymmetry has been reached. This point will be explored in more detail in 
\S\,\ref{subsec:Correlations}.

In \cite{Beniamini2020}, light curves from off-axis steep jets have been explored in detail with the aim of characterizing them using light curve features alone. These light curves exhibit one or two peaks depending on the viewing angle and jet structure. For light curves that feature two peaks, the earliest one is associated with the end of the coasting phase near our line of sight, where $\Gamma$ is constant until enough mass from the outer medium accumulates to start decelerating the blast wave. This deceleration time occurs early in the evolution of the system and is not seen in the light curves presented in this current work. The second light curve peak can be seen in all light curves presented in this section and is attributed to the geometrical effect induced by the jet core coming into view. In cases where the system geometry causes these times to be well separated, there is an additional light curve feature in the shape of a light curve dip, occurring at an observer time $t_{\text{dip}}$ which separates the end of the coasting phase from the onset of the angular structure dominated phase. During 
the latter phase, the observed emission originates from a region at an angle $\frac{1}{\Gamma}$ off the line of sight\footnote{at $\theta_F(t)$ that satisfies $\Gamma[\theta_F(t),t]=1/[\theta_{\rm obs}-\theta_F(t)]$.}, towards the jet core. Since the emission is dominated by the jet structure, the observed flux gradually rises as more energetic regions start contributing.
The transition time $t_{\rm dip}$ is found at the intersection point of two power-law curves fitted to the light curve, marked with a star and a vertical dotted line in the left panel of Fig.~\ref{fig:P2Together}.
This critical time manifests as the onset of a plateau in the polarization curve, seen most clearly in the steeper structures (in red and purple solid lines). The power-law energy structure probed at this time range causes a self-similar behavior of the emission region, resulting in constant polarization degrees which hold up to the point the observed flux switches to becoming core-dominated.
This behavior becomes less pronounced in the other setups explored in this work (corresponding to smaller values of $a$) as they present smaller dynamical range in time and do not allow enough temporal separation between $t_{\rm dip}$ and the second light curve break, $t_{\text{b}}$. Moreover, for smaller $a$ values, the effective core of the jet becomes larger (see Fig. \ref{fig:EnergyPerUnitSolidAngleExample}), allowing emission from it to reach an observer at a specific viewing angle at an earlier time compared to a steeper structure. This acts to reduce the time difference between $t_{\rm dip}$ and $t_{\rm b}$ and almost erase the polarization plateau seen in the $a=5.5$ case.
 

Another geometrical parameter that affects the observed flux and its corresponding polarization curve is the viewing angle. Its effect is demonstrated in the right panel of Fig. \ref{fig:P2Together}, where the upper panel corresponds to the polarization curve and the lower panel shows the light curves. The magnetic field structure is set to be random in the plane of the shock ($\xi\rightarrow 0$) with a steep angular energy structure ($a=3.5$). The normalized viewing angle is set to be off-axis with values of $q\ge 1$ which differ by line texture.
One notable feature is the change in the height of the polarization peak, which increases with growing values of $q$. As the observer becomes more misaligned with the jet symmetry axis the level of asymmetry in the system increases, which leads to higher levels of polarization. At the most extreme misalignment degree in our study ($q=5$, solid purple curve), the height of the polarization peak starts approaching the maximal polarization level of synchrotron radiation (e.g. \citealt{Rybicki1979,Granot2003}; B24).
As discussed for the previous plot, it can be seen that the polarization peak is associated with a break in the light curve, due to dominant contribution to the observed flux from jet core emission, which becomes visible around this time. This correlation, as well as the dependence of the polarization peak height on the normalized viewing angle, is explored in detail in 
\S\,\ref{subsec:Correlations}. The time at which the polarization peaks \footnote{and the corresponding light curve breaks or peaks} changes with the viewing angle, as it is dominated by the angular proximity of the line of sight to the jet core. The closer the observer line of sight is to the jet symmetry axis, the earlier the observed flux will be dominated by the jet core, leading to an earlier break in the light curve and a peak in the polarization curve, which correspondingly occurs at a higher flux level. A similar trend can also be seen for shallow jets (B24).

\subsection{Magnetic field 3D orientation}
\label{sec:MFOrientation}
Up until this point we investigated the effects of the geometrical parameters (the jet energy profile power-law index, $a$, and normalized viewing angle, $q$) on the observed flux and consequent polarization of GRB afterglows that involve steep jets, while assuming a random magnetic field in the plane of the shock (set with $\xi\rightarrow 0$). The effect of the magnetic field structure on the observed polarization and light curves is studied in this section by setting the viewing angle and jet angular energy structure as constants (with $q=3$ and $a=3.5$ respectively) and varying the structure of the magnetic field behind the shock with the value of $\xi$. The generally random magnetic field corresponds to an isotropic field (in 3D) that is
stretched in the radial direction (along the local shock normal) by a factor of $\xi$ \citep[following][B24]{Sari1999Pol,Gill2020}, such that $\xi\to0$ corresponds to a random field completely in the plane of the shock while $\xi\to\infty$ corresponds to an ordered field in the radial direction (for details, see B24). The polarization degree of each curve does not change sign during its temporal evolution since the viewing angle is off-axis. 

We consider five representative values of the magnetic field stretching factor
$\xi$ and calculate their corresponding polarization (Fig. \ref{fig:q>PDFAllXi}, upper panel) and light curves (Fig. \ref{fig:q>PDFAllXi}, lower panel), which are presented in shades varying from purple to gray in Fig.~\ref{fig:q>PDFAllXi}. The choice of modeled stretching parameters includes two extreme values ($\xi\rightarrow 0, \infty$), the limits imposed by the radio afterglow polarization of GW 170817 ($\xi=0.75,1.15$; \citet{GG18}, B24) and one intermediate value with $\xi>1$ as suggested by \citet{Arimoto2024}.
It is apparent from the upper panel of Fig. \ref{fig:q>PDFAllXi} that the polarization changes sign, indicating a $90^{\circ}$ rotation of the polarization vector, as the stretching factor $\xi$ crosses 1. This is caused by a change in the dominant component of the magnetic field behind the shock, from predominantly in the plane of the shock ($\xi<1$) to predominantly radial ($\xi>1$). 

For the two extreme values of the $\xi$ parameter, which correspond to the extreme magnetic field configurations: purely random field in the plane of the shock ($\xi\rightarrow 0$) and radial magnetic field ($\xi\rightarrow \infty$), we see high values of polarization at times close to a break in the light curve. Although the polarization degree does peak for the $\xi$ parameters of order unity, the height of the peak is lower due to the mixing of the two field components, such that the local polarization contributions of their emission largely cancel out, leading to low polarization. 
In the case of a completely isotropic magnetic field with $\xi=1$, the polarization degree would be $0$ throughout. 

As the value of $\xi$ grows, the polarization degree starts to exhibit a different behavior at post-peak times, where instead of reducing to zero, it asymptotes to a constant value. This can be seen in Fig. \ref{fig:q>PDFAllXi} in the case of $\xi\rightarrow \infty$ (dashed gray lines), the polarization remains constant following the break in the light curve and its asymptotic value approaches the maximal level of polarization of synchrotron radiation, $\Pi_{\rm max}$, marked by a horizontal crimson dashed line \citep{Rybicki1979}. In addition, the corresponding light curve exhibits a steeper decline post break compared to lower values of $\xi$. Such high-$\xi$ behavior is not seen when shallow jets are considered (B24). This change in asymptotic behavior can be understood through the jet structure. 

For steep jets that feature a high value of $\xi$ and are observed off-axis, the emitting region in late times is dominated by the narrow jet core. Combined with the radially dominated magnetic field, associated with high values of $\xi$, the core behaves as a point source of synchrotron radiation with an almost uniform magnetic field, which emits highly polarized radiation \citep{Rybicki1979}, leading to a constant level of polarization at these times. In the limit $\theta_{\rm obs}\ll1\ll\Gamma_c(t\obs)$, the comoving angle of the emitted photons that reach the observer relative to the radial direction, $\theta'=\arccos\mu'$, is given by
\begin{equation}
\mu'=\frac{\mu-\beta}{1-\beta\mu}\approx\frac{1-(\theta\obs\Gamma_c(t\obs))^2}{1+(\theta\obs\Gamma_c(t\obs))^2}\;.
\end{equation}
Following the break in the light curve, at times $t\obs>t_{\text{b}}$, the Lorentz factor in the core obeys
$\Gamma_c(t\obs)<1/\theta\obs$, such that $\theta'\ll1$, $\mu'\approx1-\frac{1}{2}\theta^{\prime\,2}\approx1-2(\theta\obs\Gamma_c(t\obs))^2$ and $\theta'\approx2\theta\obs\Gamma_c(t\obs)\propto\Gamma_c(t\obs)\propto t^{-\alpha_\Gamma}$.
The comoving specific emissivity scales as $j'_{\nu'}\propto(\nu')^{-\alpha}[\sin\psi']^{\epsilon}$ where $\epsilon=1+\alpha$ for optically thin  emission from an electron distribution that is isotropic in the comoving frame \citep{Granot2003}\footnote{For optically thin fast cooling power-law segments, such as F or H in the notation of \citealt{GS02}, this holds only if the electrons cooling Lorentz factor $\gamma_c$ is assumed to be independent of pitch angle, which in turn requires significant pitch angle scattering within the dynamical time. The latter is also required, however, to maintain an isotropic electron distribution in the comoving frame.},
and $\psi'$ is the pitch angle of the emitting electron, which in our case is $\psi'=\theta'$. 
Therefore, the flux density temporal decay index steepens by $\Delta\alpha_t=\alpha_\Gamma(1+\alpha)$ relative to the $\xi\rightarrow 0$ case. For PLS G, the temporal spectral index is $\alpha=\frac{p-1}{2}$ while for a uniform external medium neglecting lateral spreading $\alpha_\Gamma=\frac{3}{8}$, which leads to $\Delta\alpha_t=\frac{3(p+1)}{16}\approx0.656$ for $p=2.5$, close to the difference in slopes we get from our semi-analytical calculation. 
For $\xi>\theta_{\rm obs}^{-1}$ this steeper decay lasts until the non-relativistic transition time when $\Gamma_c(t\obs)\sim1$ and $\theta'\sim\theta=\theta\obs$. For $1<\xi<\theta_{\rm obs}^{-1}$ this steeper decay lasts until 
$\theta'\sim\xi^{-1} \Leftrightarrow 
\Gamma_c(t\obs)\sim(\xi\theta\obs)^{-1}$, i.e. up to $t\obs\sim t_b\xi^{1/\alpha_t}$, at which point the non-radial field starts dominating the emitted flux that reaches the observer and the regular flux decay rate is resumed.

\subsection{Behavior of peak polarization}
\label{subsec:Correlations}

\begin{figure}
	\includegraphics[width=\columnwidth]{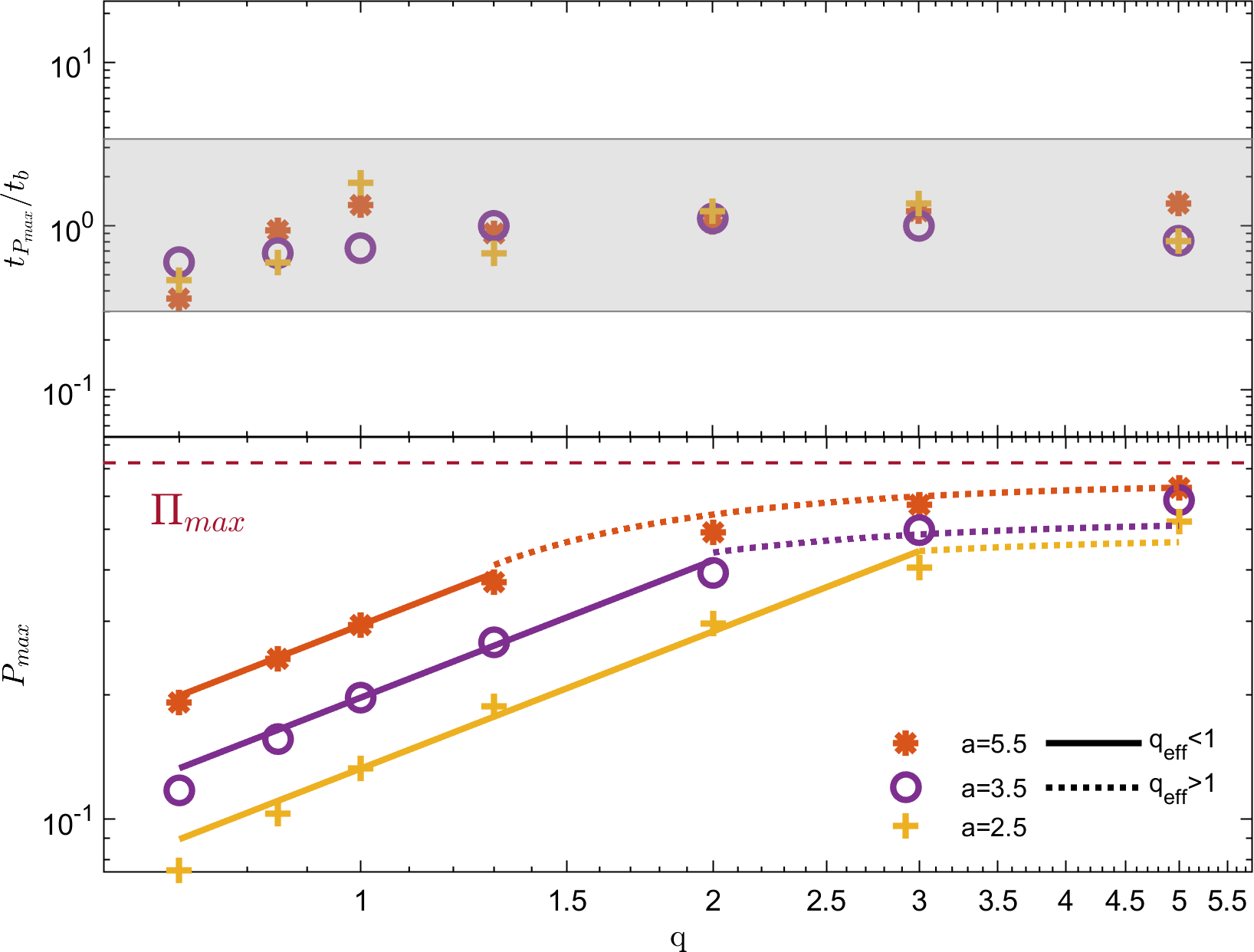}
    \caption{\textbf{\textit{Upper panel}}: Ratio between the polarization peak times $t_{\Pmax}$ and light curve break times $t_{\rm{b}}$ as function of normalized viewing angle  $q=\theta_{\rm obs}/\theta_c$ for all steep jet structures considered in this work. We can see that these two times are within a factor of three from each other. \textbf{\textit{Lower panel}}: Maximal polarization degree $\Pmax$
    as function of the normalized viewing angle $q=\theta_{\rm obs}/\theta_c$. The maximal polarization level shows an almost linear rising trend for small values of $q$ (solid lines), which correspond to systems viewed from within their effective core opening angle $\theta_{\text{c,eff}}$. This trends flattens out at larger values of $q$ (dotted lines), when the system is viewed off-axis in regard to the effective core. The size of this effective core grows as the jet becomes shallower, marking a change in the value of $q$ where this turnover occurs, $q_{\text{c,eff}}$.}
    \label{fig:1DQCorrelations}
\end{figure}

In sections \ref{subsec:OnAxisJets}, \ref{subsection:OffAxisJets} and \ref{sec:MFOrientation} we showed how the afterglow light and polarization curves change as the viewing angle, jet angular energy structure and magnetic field structure behind the shock vary. In addition, we highlighted the evident connections between these two observables.

Following previous work (\citealt{Sari1999Pol,Granot2003,Granot2003a,Rossi2004, Birenbaum2021}; B24), we highlight in the upper panel of Fig. \ref{fig:1DQCorrelations} the connection between the time the polarization level peaks ($t_{\Pmax}$) and the time a geometrical break appears in the light curve ($t_{\rm{b}}$). The magnetic field structure is assumed to be random in the plane of the shock with $\xi\rightarrow 0$.
For slightly misaligned jets, viewed from within their core opening angle, the polarization peaks when 
a narrow region of the polarized ring, upon which the polarization vector is ordered, dominates the observed emission. Following this time, the light curve experiences a steepening as the whole core has been revealed to the observer and there are no more energetic new parts of the jet that contribute to the emission. 

The off-axis structured jet scenario behaves similarly, where polarization peaks as the core is revealed to the observer and the light curve shows a break when the emission turns core-dominated. In the upper panel of Fig.~\ref{fig:1DQCorrelations}, the ratio between the two times $t_{\Pmax}/t_{\rm{b}}$ is plotted as a function of the normalized viewing angle $q$ and these two times are shown to be within factor 3 of one another (shaded gray region). This fortifies the relation between the two times and is consistent with the results shown in B24.

In the bottom panel of Fig.~\ref{fig:1DQCorrelations} we show the dependence of the peak polarization level $P_{\rm max}$ on the normalized viewing angle $q=\theta\obs/\theta_c$ for the steep jet structures considered in this work. As described in section \ref{subsection:OffAxisJets}, 
$P_{\rm max}$ rises as the viewing angle grows
due to increased asymmetry. 
In B24
we quantify this dependence using a simple toy model which uses a top hat jet and find the functional behavior changes when the system is viewed on- or off-axis. For $q<1$, the polarization level rises as $\propto q^{1.1}$ while for $q>1$ the behavior saturates into an asymptotic value. The general behavior of the polarization peak as function of $q$ can be described as $\Pmax(q)\propto\frac{\sin 2\chi}{2\chi}$ where the value of $\chi$ is set by
\begin{equation}
    \chi=   
    \begin{cases}
      \cos^{-1}\left(\frac{q}{2}\right)  & q\leq1 \;,\\
      2\sin^{-1}\left(\frac{1}{2q}\right) & q\geq1  \;.
      \label{eq:BPL}
    \end{cases}
\end{equation}
While this behavior holds for top hat jets (see B24), 
it is strongly affected by the effective core size $\theta_{\text{c,eff}}$, determined by the jet structure. This modified behavior can be seen for shallow jets ($a\leq2$), where the jet's effective core angle becomes
larger due to contributions from the energetic power-law wings,
causing systems viewed off-axis to behave as if
they are on-axis. Such systems exhibit a rising trend in peak polarization levels that behaves like $q^{1.1}$ (see B24) lasting up to a larger value of $q$ before $P_{\rm max}$ saturates.
In the bottom panel of Fig.~\ref{fig:1DQCorrelations} we see that the turnover value of $q$ at which the behavior switches becomes closer to 1 as the jet becomes steeper, indicating the effective core angle $\theta_{\text{c,eff}}$ is becoming smaller with less contribution from the jet power-law wings.
We leave the exact definition of the effective core angle for future work.

\begin{figure}
	\includegraphics[width=\columnwidth]{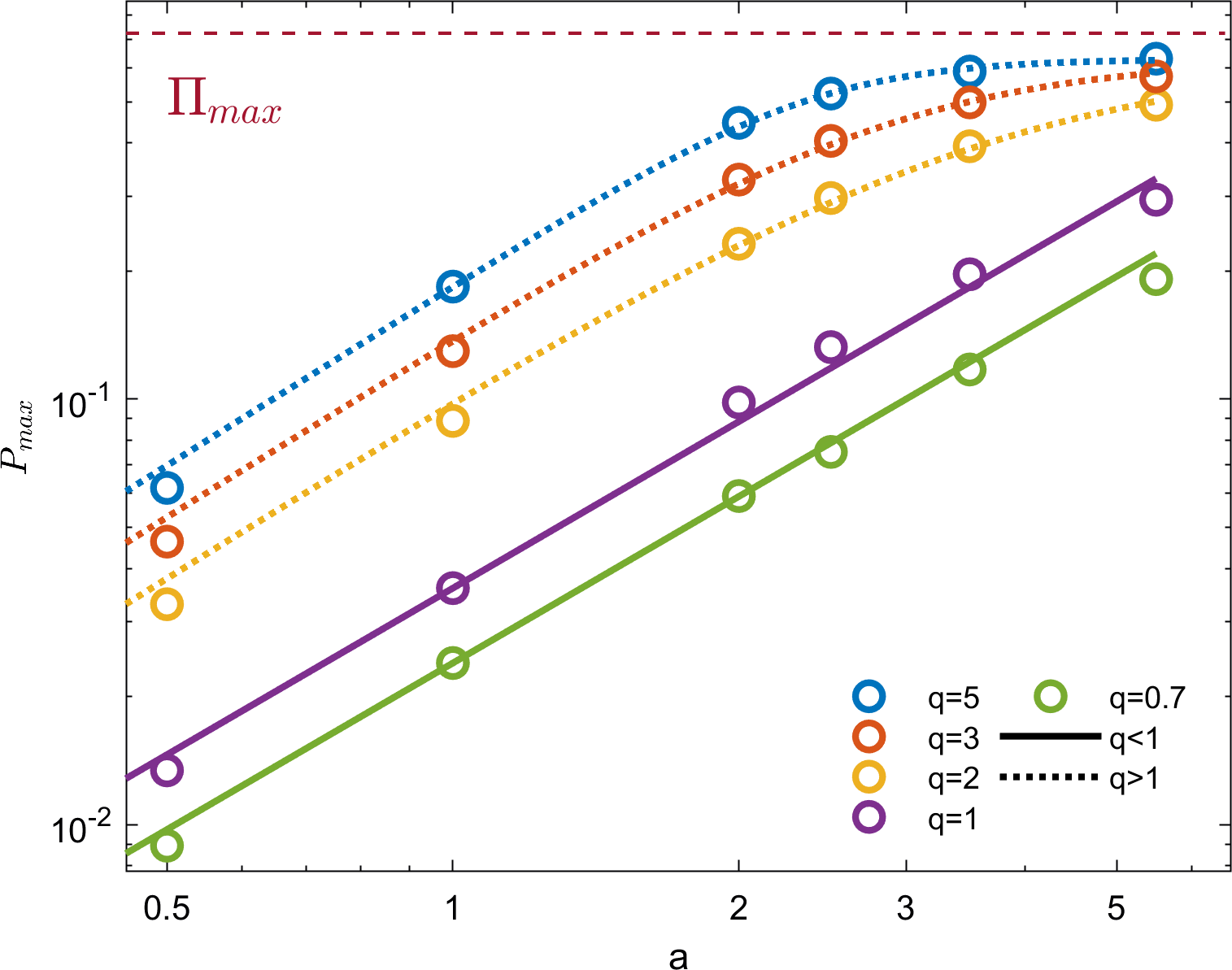}
    \caption{Maximal polarization levels, $P_{\rm max}$, as function of the power-law index of the energy angular profile, $a$, for various viewing angles and random magnetic field in the plane of the shock ($\xi\rightarrow 0$). Values of $a\leq 2$ and $a>2$ correspond to shallow and steep jets, respectively. 
    While jets
    observed from within their core opening angle $\theta_{\text{c}}$ ($q<1$, solid lines) exhibit a power-law trend, 
    $P_{\text{max}}\propto a^{1.3}$, 
    jets observed off-axis 
    ($q>1$, dotted lines) show a broken power-law behavior, which rises as $\propto a^{1.4}$ before asymptoting to a constant level of polarization. The value of $a_{\text{crit}}$ where the turnover occurs depends on $q$ due to
    the impact of the jet structure on its effective core size $\theta_{\text{c,eff}}$.}
    \label{fig:1DACorrelations}
\end{figure}

In Fig. \ref{fig:1DACorrelations} we show the correlation between the peak polarization level and the power-law index of the angular energy structure, $\Pmax(a)$, for viewing angles in the  range $0.7\leq q\leq5$. The range of $a$-values shown here expands upon the results presented in B24
and covers the parameter space of both shallow and steep jets ($0.5\leq a \leq 5.5$). We observe a different behavior with $a$ for on- and off-axis viewing angles. For on-axis observers
with $q<1$, the peak polarization level rises as
a power-law 
$\propto a^{1.3}$. This trend can be explained by the increase in the asymmetry in the system as the value of $a$ increases and the jet becomes steeper\footnote{i.e. $\theta_{\text{c,eff}}$ decreases as $a$ increases for a fixed value of $q$.}.
During the time of the peak in steeper jets, the observed flux is dominated by the energetic jet core, with a decreasing contribution from the jet power-law wings
as the jet becomes steeper, which raises asymmetry in the system and its consequent polarization. This rising behavior changes for off-axis observers
with $q>1$ where it can now be described with a broken power-law\footnote{consistent with the results of \citet{Birenbaum2024}.} that rises as $a^{1.4}$ and saturates to an asymptotic value of $\sim 0.63$ at a turnover value $a_{\text{crit}}(q)$
that depends on the viewing angle. This asymptotic behavior is a result of the system approaching the maximal polarization level possible for a random magnetic field configuration that is confined to the shock plane ($\xi\rightarrow 0$). We can describe the $q>1$ dependence with the following expression
\begin{equation}
    \Pmax(a,q,\xi\!\to\!0)=0.63\left[\left(\frac{a}{a_{\text{crit}}(q)}\right)^{-s(q)}+1\right]^{-1.4/s(q)},
\end{equation}
where the turnover value is best fit
with a power-law $a_{\text{crit}}(q)=4.84\cdot q^{-0.43}$ and the smoothness parameter of the break has been fitted with the expression $s(q)=1.14\cdot q^{0.87}$.

\begin{figure}
	\includegraphics[width=\columnwidth]{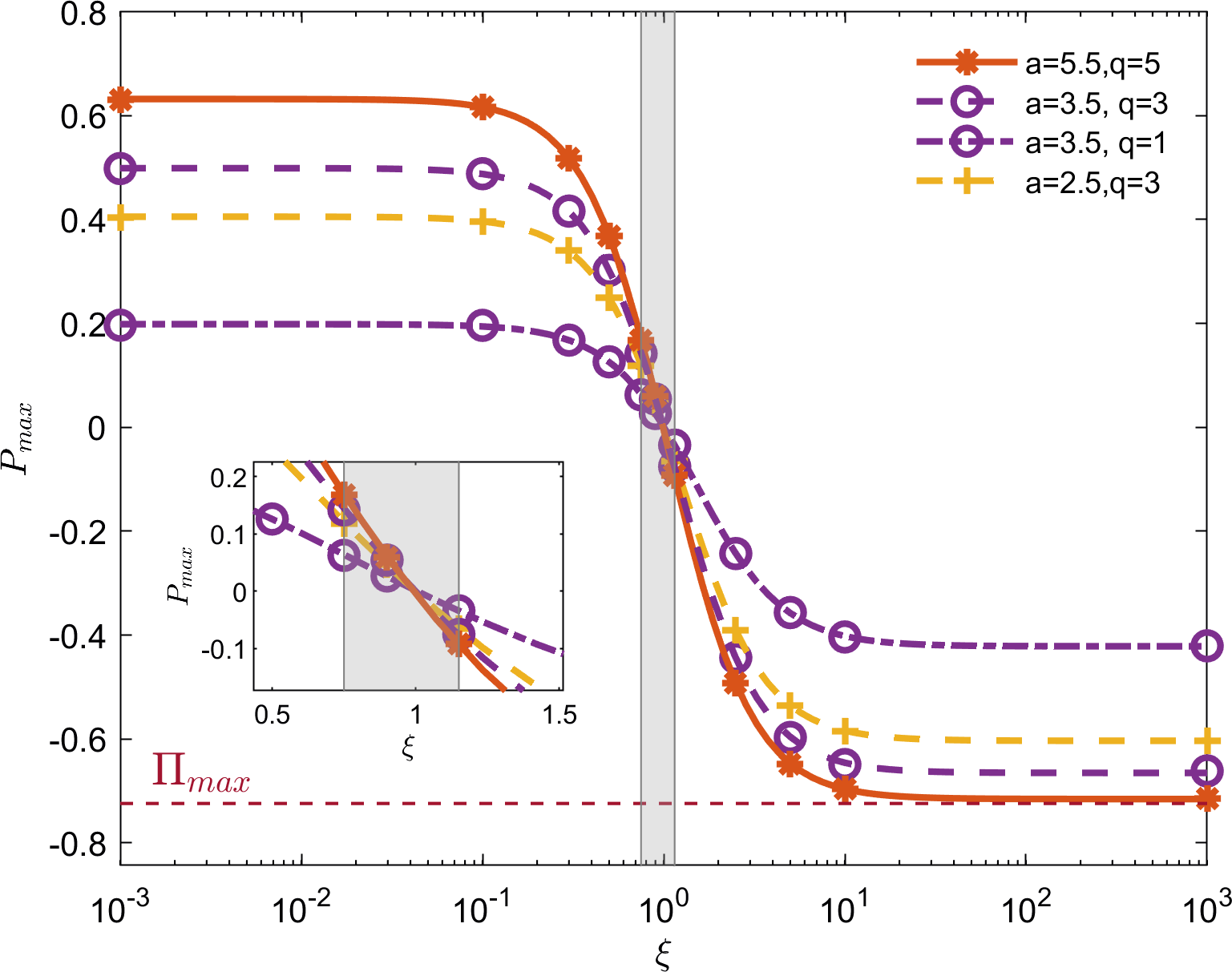}
    \caption{Maximal polarization degree $\Pmax$ as function of the magnetic field stretching factor
    $\xi$, in which synchrotron radiation is emitted. This relation is plotted for four combinations of the jet energy angular profile power-law index $a$ and normalized viewing angle $q$, and is well described by
    a hyperbolic tangent of $\log_{10}\xi$ for all values. The region of $\xi$ imposed by 
    the upper limits on the polarization of GW 170817 ($0.75<\xi<1.15$; \citealt{GG18, Gill2020}; B24) is marked in shaded gray.
    The insert zooms in on this region.
    All curves change sign at $\xi=1$. Detailed models can be found in Table~\ref{tab:maxP(xi)Table}.}
    \label{fig:1DXiCorrelations}
\end{figure}

Finally, we 
analytically study how the peak polarization degree $P_{\rm max}$ changes with the value of the magnetic field structure parameter behind the shock, by fixing the jet structure ($a$) and viewing angle ($q$) and changing the value of the stretching factor $\xi$ from $\rightarrow 0$ (random magnetic field in the plane of the shock), through intermediate values, to $\rightarrow \infty$ (radial magnetic field). 
In Fig.~\ref{fig:1DXiCorrelations} we present $P_{\rm max}(\xi)$ for several configurations of the system (combinations of $a$ and $q$). We find that $P_{\rm max}$ approaches a constant value
for each setup when the magnetic field is dominated by either the random ($\xi\ll1$) or radial ($\xi\gg1$) components. However, when these two components become comparable ($\xi\sim1$), the degree of cancellations in the system rises and reduces the observed polarization as a result. We mark in gray shaded area the limits on the magnetic field structure parameter imposed by upper limits on the polarization of the radio afterglow of GW 170817, derived by \citet{GG18}\footnote{Also derived independently by \citet{Corsi2018,Stringer2020,Teboul2021}.} and adapted to the terms of our model. This measurement points at a rather isotropic magnetic field behind the shock, which will produce low levels of polarization during the peak regardless of jet structure and viewing angle (see inset of Fig. \ref{fig:1DXiCorrelations}). When the magnetic field is completely isotropic in all 3D directions, the polarization completely vanishes. This is evident in Fig. \ref{fig:1DXiCorrelations}, with all curves crossing zero at $\xi=1$. The asymptotic polarization level of the most asymmetric geometric setup we explore in this part  (with $a=5.5$ and $q=5$ in solid red lines) has polarization levels that approach the maximal level of polarization for synchrotron radiation, $\Pi_{\text{max}}$ (crimson dashed line),
when $\xi\rightarrow\infty$. This is achieved since in this regime the emission during the polarization peak is dominated by the narrow jet core and viewed off axis, with a uniform radial magnetic field that is misdirected w.r.t the line of sight. 
The change in $\Pmax$ with the value of $\xi$ has been fitted with a hyperbolic tangent function of the form (B24): 
\begin{equation}\label{eq:pol_xi}
\Pmax(\xi)=A\tanh(C-B\log_{10}\xi)-D\;, 
\end{equation}
where $(A-D)$, and $-(A+D)$ are the asymptotic values of $\Pmax$ at $\xi\rightarrow0$ and $\xi\rightarrow\infty$, respectively, $B$ is the width of the transition region and 
$C=\tanh^{-1}(D/A)$ ensures that $\Pmax(\xi=1)=0$ as expected in a case of a completely isotropic magnetic field with $\xi=1$. The fitted expressions for the models shown in Fig. \ref{fig:1DXiCorrelations} are presented in Table~\ref{tab:maxP(xi)Table} in Appendix~\ref{app:maxP(xi)Table}, where a slight dependence of $A$, $B$, $C$ and $D$ on $a$ and $q$ can be seen.

All the descriptions for the functional dependence of the peak polarization level on the magnetic field structure (through the stretching factor $\xi$), normalized viewing angle ($q$) and angular energy structure (through the power-law index $a$) can be combined into a single expression. This expression provides an approximation for the peak polarization level given the system's geometrical conditions for values of $\xi\lesssim 1$. This expression is of the following general form
\begin{equation}
\Pmax=\Psi(a,q)\left[A\tanh(-B\log_{10}\xi+C)-D\right] ,\label{eq:GeneralAnalyticalExpression}
\end{equation}
where the values of $A$, $B$, $C$ and $D$ can be found in table \ref{tab:maxP(xi)Table} and the functional dependance on $q$ and $a$ is given by
\begin{equation}
    \Psi(a,q)=  
    \begin{cases}
      q^{1.1}\left(\frac{a}{3.5}\right)^{1.3}  & q<1 \;,\\
      \left[\left(\frac{a}{a_{\text{crit}}(q)}\right)^{-s(q)}+1\right]^{-1.4/s(q)}\frac{\sin2\chi}{2\chi}& q>1  \;,
      \label{eq:Psi(aq)}
    \end{cases}
\end{equation}
with $\chi=2\sin^{-1}\left(\frac{1}{2q}\right)$. 
We can plug in the expressions in table \ref{tab:maxP(xi)Table} to get an analytical approximation for the peak polarization level as function of system parameters. Since in this work we find different behaviors for jets viewed on- and off-axis, below we give two different suggestions that approximate the peak polarization levels of steep jets.
The expression for on-axis jets with $q<1$, $a>2$ and $\xi\lesssim 1$:
\begin{equation}
    \Pmax=   
q^{1.1}\left(\frac{a}{3.5}\right)^{1.3}\left[0.31\tanh\!\left(0.38-2.1\log_{10}\xi\right)-0.11\right]\,.
      \label{eq:AnalyticalExpqsmall1}
\end{equation}
For off-axis viewing angles, with $q>1$, $a>2$ and $\xi\lesssim 1$, we obtain:
\begin{equation}
    \Pmax=   
\Psi(a,q)\left[0.67\tanh\!\left(0.054-2.2\log_{10}\xi\right)-0.042\right],
      \label{eq:AnalyticalExpqlarge1}
\end{equation}
To re-normalize these expressions for other choices of the electron particle population power-law index $p$,\footnote{when focusing on PLS G.} which denotes the value of $\Pi_{\max}$, one can divide them by the value of $\Pi_{\max}$ and multiply by the corresponding value that fits the choice of $p$.
The expressions presented in Eqs. \ref{eq:GeneralAnalyticalExpression}-\ref{eq:AnalyticalExpqlarge1} allow us to approximate the peak polarization level using the geometrical parameters of the system. Such expressions can be used when optical afterglow polarization is observed close to the light curve break time\footnote{When the break has a geometrical origin and not a spectral one.} to constrain the geometrical properties of the system alongside modeling of the light curve, without the need for running complex models. 

\section{AT2021lfa - Orphan afterglow candidate}
\begin{figure*}
  \centering
  {\includegraphics[width=\columnwidth]{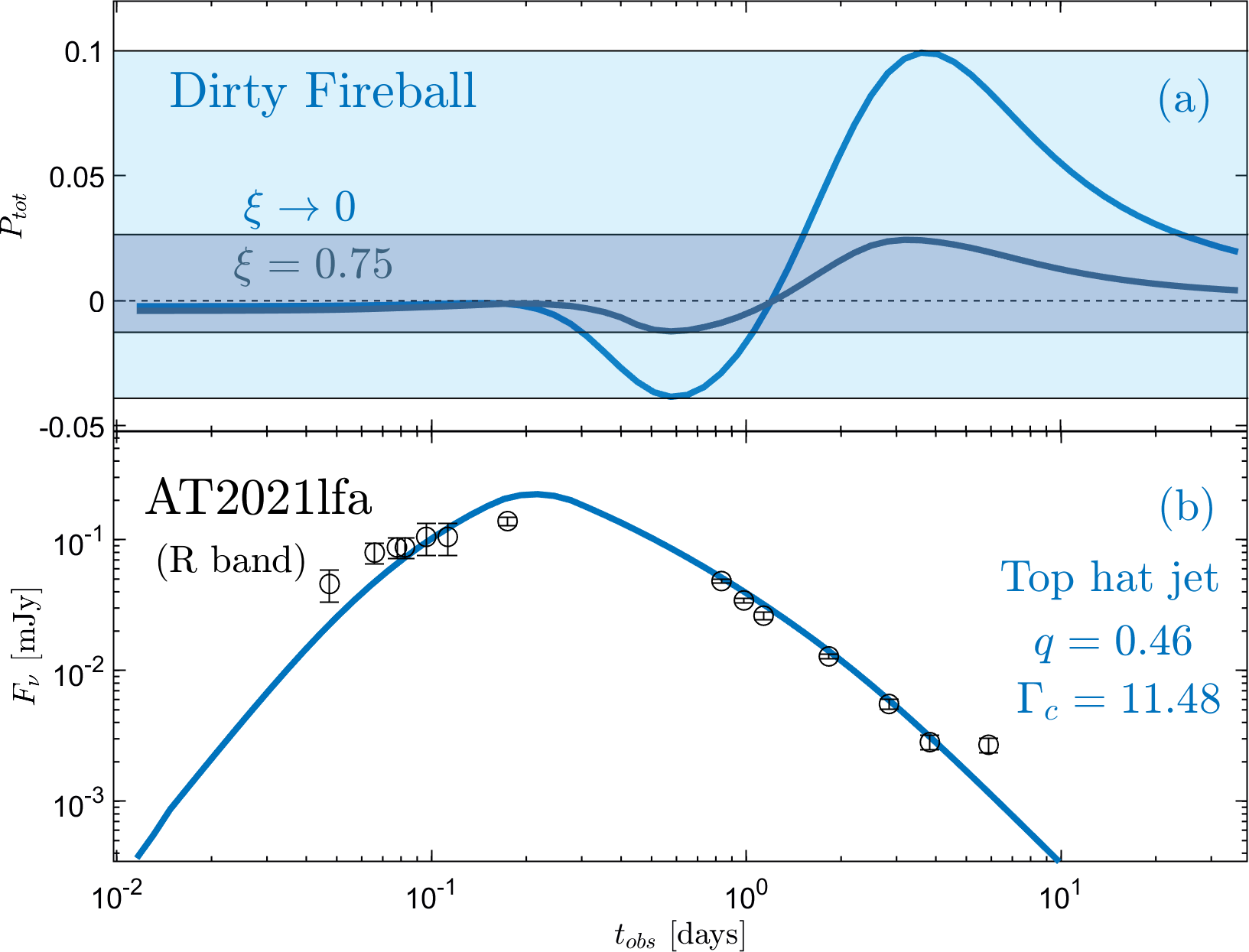}} 
  {\includegraphics[width=\columnwidth]{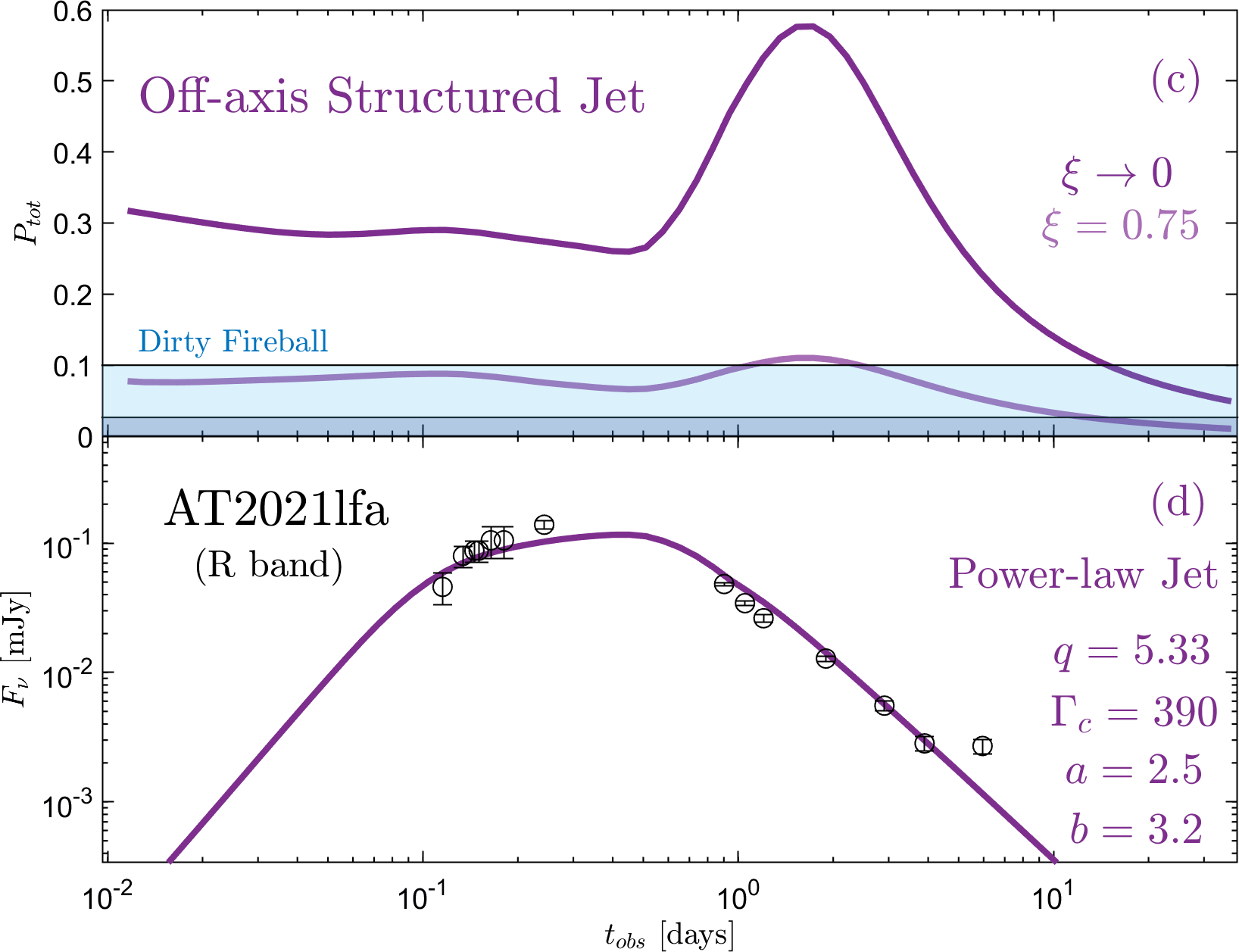}} 

  \caption{\textbf{\textit{Upper panels}}: observed polarization curves corresponding to the dirty fireball (panel (a)) and off-axis structured jet (panel (c)) models for the R-band (optical) light curve 
  of AT2021lfa. 
  Polarization curves are presented for two values for the magnetic field stretching factor $\xi$: 
 a random magnetic field confined to the plane of the shock ($\xi\rightarrow 0$) and the more realistic $\xi=0.75$ which also has a slightly weaker radial component. 
 The $\xi\rightarrow 0$ models exhibit higher polarization than the $\xi=0.75$ models. The peak levels of polarization of the dirty fireball scenario are marked with blue (for $\xi\rightarrow 0$) and dark blue (for $\xi=0.75$) shading, 
 for comparison with the off-axis structured jet scenario. 
  \textbf{\textit{Lower panels}}: observed R-band flux for AT2021lfa (black data points, taken from MASTER-OAFA and ZTF; \citealt{Lipunov2021,Lipunov2022, Yao2021a}),
  and computed light curves according to the dirty fireball (panel (b), solid blue lines) and off-axis structured jet (panel (d), solid purple lines). Both models are able to describe the trends seen in the flux observations while having very different geometrical parameters. Detailed afterglow parameters can be found in Table~\ref{tab:AT2021lfaParameterSpace}.
  }
 
\label{fig:AT2021lfaXi0AndXi075Lim}
\end{figure*}

\begin{table}[h!]
 \caption{Afterglow model parameters for the dirty fireball and off-axis structured jet models of the R band flux observations of AT2021lfa. The produced R band light and polarization curves are presented in Fig \ref{fig:AT2021lfaXi0AndXi075Lim}. The dirty fireball model is adapted from \citet{Li2024}.}\label{tab:AT2021lfaParameterSpace}
\centering
\renewcommand{\arraystretch}{1.4} 
\begin{tabular}{l c c c}
\hline\hline
 Parameter & Dirty Fireball & Off-axis 
 Jet\\
 \hline
 $\theta_c$    &$13.2^{\circ}$ & $1.72^{\circ}$\\
  $\theta\obs$    &$6.3^{\circ}$ &$9.2^{\circ}$\\
 $\Gamma_c$  & 11.5 & 390\\
 $a=-\frac{d\log E_{\rm k,iso}}{d\log\theta}|_{\theta>\theta_c}$ 
 &  0  &2.5\\
$b=-\frac{d\log(\Gamma_0-1)}{d\log\theta}|_{\theta>\theta_c}$ 
&0 &3.2 \\
 $E_{\text{c}}$ [erg]& $9.2\cdot 10^{ 50}$ &  $3.5\cdot 10^{52}$ \\
 $n_{\text{ISM}}$ [cm$^{-3}$]&  49.7 &  70\\
 $k=-\frac{d\log \rho}{d\log r}$
 & 0 
 & 0 
 \\
  $p$&  $2.53$ & $2.37$\\
 $\epsilon_\text{e}$   & $10^{-0.32}$ & 0.17 \\
 $\epsilon_\text{B}$&  $10^{-1.62}$ & 0.3\\
  $\chi_\text{e}$&  0.48 & 1\\
$T_0$ [days] 
& 59338.06 & 59338.01\\
(estimated GRB time) & & \\
\hline
\end{tabular}
\end{table}

The optical transient AT2021lfa was first detected by ZTF at 05:34:48 UTC 2021 May 4. Followup observations found a power-law like flux decay, with no preceding detections and a source
redshift of $z=1.063$ \citep{Yao2021b,Yao2021a}. Three hours before the first ZTF detection, the MASTER-OAFA robotic telescope recorded the same transient during a routine sky survey, showing a rising trend in observed flux \citep{Lipunov2021,Lipunov2022}. This transient has also been detected in radio, however the observations are suspected to heavily suffer from interstellar scintillation \citep{Li2024}.
Comparison between the R-band light curve and other optical afterglows shows similarities which hint at a relation between this optical transient and GRBs \citep{Lipunov2022}. Additional \textit{Swift}-XRT observations demonstrate a typical photon index that can be related to GRBs as well as brightness that is similar that of X-ray afterglows. Using correlations between the X-ray flux and prompt $\gamma$-ray brightness, \citet{Lipunov2022} deduce that if indeed this transient is related to a GRB, its prompt $\gamma$-ray emission should have been observed. However, searches in the archives of space $\gamma$-ray observatories did not yield a counterpart. 

Afterglow-like transients that lack an associated prompt $\gamma$-ray emission are termed ``orphan afterglow candidates".
If these transients are indeed related to GRBs, the lack of observed prompt $\gamma$-ray counterpart can be attributed to one of three reasons \citep{Rhoads2003,Huang2002,Lipunov2022,Li2024}:
\begin{enumerate}
    \item The GRB system was observed beyond the relativistic jet core, causing the prompt emission that comes from the jet core to be missed (while outflow along our line of sight is not relativistic enough to efficiently produce $\gamma$-rays, see \citealt{2019MNRAS.482.5430B}). A rising phase in the afterglow light curve can be attributed to emission from matter in the jet wings along our
    line of sight \citep{GG18, Beniamini2020}. Such systems are termed in this work as \textbf{off-axis structured jets}.
    \item The jetted matter did not 
    reach a high enough Lorentz factor to efficiently produce prompt $\gamma$-ray emission due to baryon entrainment in the ejecta, which slows it down. In this scenario, our line of sight 
    can be within the jet's core. 
    Although such a system will not produce a prompt $\gamma$-ray signal, it can produce an observable afterglow. We refer to such systems as \textbf{dirty fireballs}.
    \item The prompt $\gamma$-ray emission was not observed due to incomplete sky coverage.
\end{enumerate}

It is difficult to 
observationally distinguish between the
different scenarios for orphan afterglows, whose prompt emission was not detected for intrinsic physical reasons (1 and 2 above). \citet{Huang2002} and \citet{Rhoads2003} discuss ways to distinguish between the different scenarios based on their light curve evolution, as well as testing the long-term radio evolution.

The light curve analysis approach is taken by \citet{Lipunov2022} and the authors fit the optical light curve with a top hat jet, viewed on- and off-axis. The authors conclude that if AT2021lfa is indeed an orphan afterglow, it must be the result of a dirty fireball, as the on-axis scenario manages to reproduce the rising phase of the optical light curve better than the off-axis one. They estimate an initial Lorentz factor of $\sim 20$. The same approach is taken by \cite{Li2024} and the authors use the \texttt{afterglowpy} tool, which also takes into account the possibility of a structured jet, and use it to fit different models to the light curve. They find that the optical light curve is consistent with both the dirty fireball and the off-axis structured jet scenarios. The radio observations greatly suffer from interstellar scintillation, making them hard to model. While such an approach is exciting, it also emphasizes that light curve modeling alone sometimes cannot distinguish between  different scenarios for orphan afterglow candidates.

In addition to the methods mentioned above \citep{Huang2002, Rhoads2003}, \citet{Granot2002} suggested that polarization measurements should be included in orphan afterglow searches in order to constrain the GRB jet opening angle distribution. Here we follow these suggestions and
demonstrate how strategically measured linear optical polarization can assist in discerning between the two intrinsic scenarios for the lack of observed prompt $\gamma$-ray emission for orphan afterglows. Our main result is presented in Fig. \ref{fig:AT2021lfaXi0AndXi075Lim}, in which we present afterglow models that follow the trends seen in the observed R-band light curve of AT2021lfa and show vastly different polarization signatures. 

The \textbf{dirty fireball} scenario for the R-band observations of AT2021lfa is presented in the left hand side of Fig. \ref{fig:AT2021lfaXi0AndXi075Lim}. Panel (b) shows the light curve, based on an adapted version of the fitted top hat jet model of \citet{Li2024}. The details of the afterglow model parameters are presented in Table~\ref{tab:AT2021lfaParameterSpace}. The main features of the model are its low initial Lorentz factor ($\Gamma_c=11.48$) that does not enable prompt $\gamma$-ray production and emission, which means it won't be observed even though the system is on-axis with $q=0.46$ \citep{Huang2002, Rhoads2003}. While the main features of the \citet{Li2024} dirty fireball  model were kept in our analysis (system geometry and initial Lorentz factor), some adaptations were made to the afterglow model to account for flux normalization differences between our code and the \texttt{afterglowpy} tool. At early times, the modeled light curve shows a rising trend, which is meant to explain the MASTER observations. This trend corresponds to the pre-deceleration phase of the light curve, at which the flow coasts at a constant $\Gamma=\Gamma_0$, culminating in a deceleration peak, as the dynamical solution transitions to the Blandford-Mckee deceleration phase \citep{1976PhFl...19.1130B,GG18}. 
Following this deceleration time, the light curve shows a declining power-law evolution that follows the trend of the observations, with a slight steepening at $\sim 2$ days that corresponds to the jet break. 

Panel (a) of Fig.~\ref{fig:AT2021lfaXi0AndXi075Lim} shows the corresponding polarization curves for
two values of the magnetic field stretching factor: 
$\xi\rightarrow 0$, which represents a purely random magnetic field in the plane of the shock, and the more realistic $\xi=0.75$ (in blue and dark blue solid lines, respectively). In the pre-deceleration phase, the observed polarization degree for both models remains constant with a 
value $\rightarrow 0$. Since the flow is coasting at constant velocity, the initial angular scale of the emitting region on the shock face does not change with time ($1/\Gamma_0$),
which leads to constant observed polarization levels\footnote{that depend on the initial geometrical setup of the system.}. As the dynamical solution changes from coasting to Blandford-Mckee following the deceleration peak, edge effects begin to affect the observed region and the polarization signature of a top hat jet, observed on-axis, appears. This temporal evolution is composed of two polarization peaks at opposing signs which correspond to a $90^\circ$ difference in the polarization angle.
In the slow cooling regime, where $\nu_{\text{m}}<\nu\obs<\nu_\text{c}$, the 
afterglow image is limb-brightened \citep{Sari1998Egg,Granot-Loeb2001,Granot2008} and may be approximated as
a bright ring around the line of sight with an angular radius of $1/\Gamma$ upon which the polarization vector is radially oriented for $\xi<1$
\citep{Sari1999Pol,Granot2003a,Nava2015a,Shimoda2020,Birenbaum2021}. When this ring is fully visible, the observed polarization is zero as cancellations across the ring are maximal in this symmetric system. However, as the shock decelerates, the 
visible region grows and eventually parts of the emitting ring disappear beyond the sharp jet edge, making the observed system less symmetrical around the line of sight. The first polarization peak occurs when a quarter of the ring is beyond the jet edge, which leads to negative polarization values, opposite in sign to the polarization values seen in previous section. 
When only half of the polarized ring remains visible, the polarization vanishes (as there is an instantaneous full cancellation) and then reappears rotated by $90^\circ$. The polarization degree reaches its second peak, with positive values, close to the jet break time as only a quarter of the polarized ring dominates the observed region (see \citealt{Birenbaum2021} for a visualization of this evolution).
While the polarization signature is similar, its magnitude changes according to the structure of the magnetic field behind the shock. The random magnetic field structure, confined to the plane of the shock ($\xi\rightarrow 0$, blue solid line) demonstrates polarization levels that reach $10\%$ at peak (shaded blue region), while the more realistic value of $\xi=0.75$, with a sub-dominant but comparable radial component,
shows peak polarization level of $\sim\!2.5\%$ (shaded dark blue region). 

The \textbf{off-axis structured jet} scenario is shown in the right part of Fig. \ref{fig:AT2021lfaXi0AndXi075Lim} where panel (d) shows the observed R-band flux and a corresponding light curve, while 
panel (c) shows the polarization curves for two values of the magnetic field 
stretching factor:
$\xi\rightarrow 0$, and $\xi=0.75$ (in purple and light purple solid lines, respectively). The main features of this model are a high initial Lorentz factor at the core ($\Gamma_{\rm{c}}=390$) that, when combined with 
a large off-axis viewing angle ($q=5.33$), can explain the absence of an observed prompt $\gamma$-ray emission component. 
A steep angular structure beyond the jet core, in both energy ($a=2.5$) and initial Lorentz factor ($b=3.2$), allows for a shallower rise in flux that better explains the observed R-band flux at early times \citep{Beniamini2020, Teboul2021, Li2024}. In the absence of spectral information, this model was constructed based on the observational optical features of AT2021lfa using the closure relations from \citet{Nakar2002} and \citet{Beniamini2020}.

In the off-axis structured jet scenario, the rise in optical flux is attributed to the angular structure dominated phase (see explanation in 
\S\,\ref{subsection:OffAxisJets} and in \citealt{Beniamini2020}). The light curve features a flat, prolonged peak due to an additional spectral crossing of $\nu_{\text{c}}$ by the observed frequency $\nu\obs$ during the geometrical peak time. The declining observed flux is attributed to the core-dominated phase, where the most energetic part of the jet has already been unveiled and is decelerating (see \S\,\ref{subsection:OffAxisJets} and \citealt{Beniamini2020}). The matching polarization curves show an almost constant polarization degree during the rising phase of the light curve, as the flux is dominated by the power-law wings of 
the energy distribution, with a corresponding power-law brightness distribution, resulting in a
self-similar behavior in terms of polarization. The level of polarization is determined by the combination of the offset from the jet symmetry axis and outflow geometry (i.e. jet angular structure), and for off-axis observers it
can reach moderate to high levels, depending on the magnetic field structure (see Fig.~\ref{fig:q>PDFAllXi} and related explanations in \S\,\ref{subsection:OffAxisJets}). As the observed emission turns core-dominated, the polarization levels begin to rise, culminating in a peak whose height depends on the geometrical parameters of the system (see \S\,\ref{subsec:Correlations}). The polarization degree remains at a constant sign due to the off-axis orientation of the observer (see \S\,\ref{subsection:OffAxisJets} and B24). 

Comparison between the left and right panels of Fig.~\ref{fig:AT2021lfaXi0AndXi075Lim} demonstrates the potential of polarization measurements for differentiating between the dirty fireball and off-axis structured jet scenarios for orphan afterglow candidates. When we consider the same magnetic field structure behind the shock for the two scenarios, stark differences arise between the dirty fireball and off-axis structured jet models in the polarized regime. For both values of the magnetic field structure parameter $\xi$, there's a factor $5-6$ difference in polarization peak height. For $\xi\rightarrow 0$, the off-axis structured jet model reaches a polarization peak of $57\%$ close to the onset of the declining phase of the light curve while the dirty fireball model peaks later with a polarization peak height of $10\%$. Similar differences can be seen when considering the more realistic value of $\xi= 0.75$ with $11\%$ and $2.5\%$ respectively.

It is also evident that the off-axis structured jet model polarization remains above that of the dirty fireball at all times when considering the same value for the $\xi$ parameter for both models. Another feature that can be seen in the dirty fireball polarization curves (due to the assumption of a top-hat jet) is the presence of a $90^{\circ}$ rotation of the polarization angle, shown as a sign change of the polarization level. Such a feature is not seen in the polarization curve of the off-axis structured jet model and can possibly be another distinguishing factor between these two models.

While comparison between models that assume the same magnetic field structure behind the shock shows clear differences between the possible interpretations for orphan afterglows, one must remember that this parameter is still not well constrained by observations and it is not clear whether it varies between bursts. This can complicate the interpretation of sparse polarization measurements for orphan afterglows. For example, if we compare between the polarization levels of the off-axis structured jet scenario with $\xi=0.75$ (Fig.~\ref{fig:AT2021lfaXi0AndXi075Lim} , light purple solid lines) and those of the dirty fireball model with $\xi\rightarrow 0$ (Fig.~\ref{fig:AT2021lfaXi0AndXi075Lim} , blue solid lines), measured about 2-3 days after $T_0$, we will get similar levels of polarization from both models ($\sim 10\%$). However, given that extreme values of $\xi\ll1$ or $\xi\gg1$ can be safely ruled out by current afterglow observations \citep{Granot2003a,GG18, Gill2020, Stringer2020}, measuring $P_{tot}\approx10\%$ would strongly favor an off-axis jet model over a dirty fireball model.

The results shown in this section clearly demonstrate the potential benefit of strategically measuring polarization for orphan afterglow candidates. Not only can such detection confirm the synchrotron source of the emission \citep[e.g.][]{Gruzinov1999,Ghisellini1999,Sari1999Pol}, but combining this with detailed light curve modeling can shed light on the intrinsic reason the prompt $\gamma$-ray emission was not observed.

\section{Discussion \& Conclusions}
\label{sec:Discussion}

In this work we explore the polarization signature of afterglows from steep GRB jets and its relation to the observed flux density.
We demonstrate how measured afterglow polarization, alongside detailed modeling of its light curve, can provide crucial information regarding the jet structure, which is shaped by the processes the jet underwent before breaking out of its confining medium. This geometrical information can be crucial when trying to decipher the origin of orphan afterglow candidates. 
While light curve based models can be degenerate, we show the expected polarization can differ, providing motivation to measure polarization for such events.

Using the semi-analytical tool developed in B24, which 
assumes an axis-symmetric 2D shock, we focus on power-law jet structures and expand the results to the regime where most of the energy is concentrated in the narrow jet core. Such models are motivated by both numerical simulations of GRB jets breaking out of their confining media and by observations of recent GRBs. We find that for a fixed viewing angle, while light curves may remain similar, the polarization levels greatly change when the jet angular structure 
is varied. Generally, the steeper the structure, the greater the observed polarization levels become due to increased asymmetry of the (unresolved) afterglow image. The polarization levels also depend on the viewing angle to the system 
and on the magnetic field structure behind the shock, where larger viewing angles and extreme magnetic field structures 
translate to higher polarization values.

It is unclear
whether the structure of the magnetic field behind the shock is similar for all GRB afterglows or not. In our model, the multi-wavelength afterglow is the result of synchrotron emission from the same particle population ("one-zone" model).
Using this assumption, we can apply the constrained range of values for the magnetic field 
stretching factor $\xi$
inferred from the upper limits on the radio afterglow polarization of GW 170817 to the optical band \citep{GG18,Corsi2018,Gill2020,Stringer2020,Teboul2021}. When applied for various steep jet geometries, the peak polarization level is less than $17\%$
(see inset in Fig. \ref{fig:1DXiCorrelations}). Similar conclusions were drawn by considering a large sample of GRB polarization measurements \citep[e.g.][]{Granot2003a,Stringer2020}. If this range of values, $0.75<\xi<1.15$, corresponding to a field rather close to isotropic, indeed represents the entire population of GRB afterglow forward shocks, we should expect relatively low levels of polarization (of $\lesssim 17\%$) even in the most asymmetric geometrical setups with off-axis viewing angles and steep jet structures.

Joint analysis of the polarization and light curves 
shows temporal proximity of the polarization peak and light curve break. These two phenomena are related to the deceleration of the jet core and the consequent revelation of the most energetic part of the system to the observer. While this connection is referred to in previous works \citep{Ghisellini1999,Sari1999Pol,Granot2002, Granot2003, Granot2003a, Rossi2004, Birenbaum2021}, in this work we quantify and visualize this relation by plotting the ratio between the polarization peak and light curve break times. 
We find that these times are within factor 3 of one another, extending the similar result of B24 
to the regime of steep jets. The important implication of this relation is that polarization measurements conducted close to the light curve geometrical break are more likely to reflect the system's peak polarization,
allowing for strategic planning of polarization measurement epochs. 

Using our model we find the peak polarization level depends on the jet structure, viewing angle and magnetic field structure behind the shock. Generally, the level of polarization rises as the jet becomes steeper and the viewing angle increases. For large off-axis viewing angles, the polarization begins to saturate up to an asymptotic value that depends on the structure of the magnetic field behind the shock. In the extreme case of a radial magnetic field, this value will be $\Pi_{\max}$ \citep{Rybicki1979, Granot2003,Birenbaum2021}. When combining the dependence of the polarization peak on the various geometrical system parameters, we find analytical approximations for the polarization peak of the form $\Pmax=\Psi(a,q)\left[A\tanh(-B\log_{10}\xi+C)-D\right]$, where the functional form of $\Psi(a,q)$ and the values of $A$, $B$, $C$ and $D$ depend on whether the viewing angle is on- or off-axis (see Eqs. 
~(\ref{eq:GeneralAnalyticalExpression})\,--\,(\ref{eq:AnalyticalExpqlarge1}) and appendix \ref{app:maxP(xi)Table}). These expressions can be useful when polarization is measured close to the light curve break time, and the jet structure and viewing angle are constrained by light curve fits, allowing for an estimation of the magnetic field structure behind the shock without the need to run complex models. 

The framework developed in this work can be useful for understanding whether 
orphan afterglow candidates (e.g. detected by ZTF or by the upcoming Vera Rubin observatory) originate from dirty fireballs, featuring low Lorentz factors and on-axis viewing angles, or are they a product of a structured off-axis jet. In addition, detecting polarization in these systems can help tie them to GRBs \citep[e.g.][]{Gruzinov1999,Ghisellini1999,Sari1999Pol}.
We demonstrate the potential of strategic polarization measurements on the observed R-band light curve of AT2021lfa that features as rising phase which turns into a power-law decline \citep{Yao2021a, Lipunov2021, Lipunov2022}. Efforts to interpret this light curve show matches with both the dirty fireball and structured off-axis jet scenarios \citep{Li2024}. However, modeling the optical polarization along with the light curve shows significant differences between these two scenarios, demonstrating the ability of such measurements to uplift the light curve degeneracy of afterglow models. While the magnetic field structure behind the afterglow shock or its universality are not well known,
there are some observational indications for it being close to isotropic. We evaluate the corresponding polarization of the two scenarios for AT2021lfa for such nearly isotropic magnetic field structure and find that while the light curves are similar, the polarization signature is different 
in several respects:
\begin{enumerate}
    \item During the rising phase of the optical light curve, the off-axis structured jet scenario shows non-zero constant polarization levels for $\xi\neq 1$ while the dirty fireball model has very small polarization levels  
    for all magnetic field structures.
    \item The heights of the main polarization peak differ by factor $\sim 5$, where the polarization of the off-axis structured jet model is higher 
    ($\sim\,$11\%) 
    than that of the dirty fireball model ($\sim\,$2.5\%). 
    \item The polarization angle changes  
    by $90^{\circ}$ (as the polarization degree temporarily vanishes) in 
    the dirty fireball scenario, while it remains constant for the off-axis structured jet scenario.
    
\end{enumerate}
Each one of the polarization signature features listed above is attributed to a different time during the evolution of the system, and while the first two features can be identified using single 
observations that are tied to specific light curve features, detecting the polarization angle change
would require extended monitoring of the system. These conclusions also hold when comparing these same models while assuming the more extreme magnetic field structure -- random in the shock plane, albeit different values for the height of the polarization peaks.

The findings listed above highlight the potential of measured polarization in discerning between the different theoretical explanations for the intrinsic absence of a prompt $\gamma$-ray emission component for observed orphan afterglow transients. We urge the community to make an attempt to measure the polarization of these transients along with their light curve, so that a combined modeling effort can discern between the different scenarios for the origin of these systems.

\begin{acknowledgements}
We thank Maggie Li for providing the observational data. We also thank Vikas Chand and Minhajur Rahaman for helpful advice.
PB and GB are supported by a grant (no. 1649/23) from the Israel Science Foundation.
PB is also supported by a grant (no. 2020747) from the United States-Israel Binational Science Foundation (BSF), Jerusalem, Israel and by a grant (no. 80NSSC 24K0770) from the NASA astrophysics theory program.
\end{acknowledgements}
\bibliographystyle{aa}
\bibliography{aanda}
 \begin{appendix}
\section{Functional dependence of peak polarization on $\xi$}
\label{app:maxP(xi)Table}
The analytical expression that relates the peak polarization degree to the magnetic field structure parameter $\xi$ has a hyperbolic tangent functional form  (Eq. \ref{eq:pol_xi}). In table \ref{tab:maxP(xi)Table} we present the functional fits plotted in Fig. \ref{fig:1DXiCorrelations} for different parameter combinations. 

\begin{table}[h!]
    \caption{Analytical functional dependence of the polarization peak height for the different models considered in Fig. \ref{fig:1DXiCorrelations}.}
    \label{tab:maxP(xi)Table}
\centering
\renewcommand{\arraystretch}{1.4} 
\begin{tabular}{l c c c}
\hline\hline
 Parameters & Functional dependence \\
 \hline
   $a=5.5$, $q=5$  &   $0.67\tanh\left(-2.2\log_{10}\xi+0.054\right)-0.042$\\
 $a=3.5$, $q=3$  &   $0.58\tanh\left(-2.2\log_{10}\xi+0.14\right)-0.083$\\
 $a=3.5$, $q=1$  &   $0.31\tanh\left(-2.1\log_{10}\xi+0.38\right)-0.11$\\
 $a=2.5$, $q=3$  &   $0.5\tanh\left(-2.2\log_{10}\xi+0.2\right)-0.1$\\
\hline
\end{tabular}
\end{table}

 \end{appendix}
%
%

\end{document}